\documentclass[12pt]{iopart}
\usepackage[utf8]{inputenc}
\usepackage{color,graphicx}
\usepackage[margin=1.0in]{geometry}
\usepackage{changepage}
\usepackage[ruled, vlined]{algorithm2e}
\usepackage{amsfonts}
\usepackage{bbm}
\usepackage{iopams}
\usepackage{amsthm}

\expandafter\let\csname equation*\endcsname\relax

\expandafter\let\csname endequation*\endcsname\relax

\usepackage{amsmath}
\usepackage{cite}
\usepackage{url}

\providecommand{\keywords}[1]{\textbf{\textit{Keywords: }} #1}

\theoremstyle{definition}

\theoremstyle{remark}
\newtheorem*{remark}{Remark}

\usepackage{subcaption}
\usepackage{array,multirow}

\begin{document}

\title[Deep learning for source detection]{Deep learning for $2D$ passive source detection in presence of complex cargo}
\today
\author{W. Baines$^1$, P. Kuchment$^2$, and J. Ragusa$^3$}

\address{$^1$ Mathematics Department, Texas A\&M University,
College Station, TX, USA}
\address{$^2$ Mathematics Department, Texas A\&M University,
College Station, TX, USA}
\address{$^3$ Nuclear Engineering Department, Texas A\&M University,
College Station, TX, USA}
\begin{abstract}
Methods for source detection in high noise environments are important for single-photon emission computed tomography (SPECT) medical imaging and especially crucial for homeland security applications, which is our main interest. In the latter case, one deals with passively detecting the presence of low emission nuclear sources with significant background noise (with Signal To Noise Ratio ($SNR$) $1\%$ or less). In  passive emission problems, direction sensitive detectors are needed, to match the dimensionalities of the image and the data. Collimation, used for that purpose in standard Anger $\gamma$-cameras, is not an option. Instead, Compton $\gamma$-cameras (and their analogs for other types of radiation) can be utilized. Backprojection methods suggested before by two of the authors and their collaborators enable detection in the presence of a random uniform background. In most practical applications, however, cargo packing in shipping containers and trucks creates regions of strong absorption and scattering, while leaving some streaming gaps open. In such cases backprojection methods prove ineffective and lose their detection ability. Nonetheless, visual perception of the backprojection pictures suggested that some indications of presence of a source might still be in the data. To learn such features (if they do exist), a deep neural network approach is implemented in 2D, which indeed exhibits higher sensitivity and specificity than the backprojection techniques in a low scattering case and works well when presence of complex cargo makes backprojection fail completely.
\end{abstract}
\keywords{source detection, Compton camera, illicit nuclear material}

\date{\today}
\maketitle
\section{Introduction} \label{S:Introduction}
Checking for presence of illicit nuclear materials (most probably in small quantities and shielded by cargo) at border crossings and shipping cargo containers in harbors is an important homeland security task. Ideally, one would try to reconstruct from the detected signals the source distribution inside the cargo. When the data is sufficiently well behaved (e.g., in SPECT), analytic reconstruction is often possible \cite{TKK}. However, in a very low SNR environment, as in the case of illicit nuclear source detection, this is impossible. Indeed, the forward analytic (integral transform type) models are not applicable. Moreover, even if they were, attempts of any filtration in FBP-type techniques lead to disaster. The saving grace is that in this case practitioners are mostly interested in getting reliable (i.e., with low rates of false positives and false negatives) information about the presence of a source, rather than its exact location.

\begin{figure}[ht!]
\centering
\includegraphics[width=0.25\textwidth]{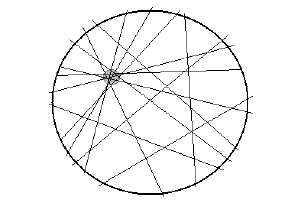}
\caption{An idea of the backprojection method.} \label{F:BP}
\end{figure}

As is well known, in passive emission imaging detectors must be direction sensitive. Indeed, otherwise the data measured has insufficient dimension for recovery of an image. Directional information is especially critical when SNR is too low for the  intensity fluctuations that arise due to the presence of a source to be statistically significant. The following options for obtaining directional sensitivity are available:
\begin{itemize}
  \item\emph{ Mechanical collimation}, when only rays incident along (or close to) a certain line are allowed to reach the detector (see Section \ref{S:compton}). This, while determining the incoming photon's direction, significantly reduces the signal strength and thus becomes unsuitable for low SNR.
  \item \emph{Compton $\gamma$-cameras} represent a more recent, and gaining its appreciation, type of $\gamma$ radiation detectors that determine a surface cone of possible incident trajectories, rather than the exact directions.
  \item Neutron detectors are being developed that (albeit based on different physics principles) produce similar cone information and lead to similar mathematical analysis.
\end{itemize}
 Backprojection detection technique introduced in \cite{ADHKK,X} relied upon finding suspicious \textbf{locations}. It utilized the following three assumptions:
 \begin{enumerate}
 \item geometric smallness of the source (usually of linear dimension on the order of $1\%$ of the linear cargo size);
 \item existence of a sufficient number of particles from the source reaching the detector being \textbf{ballistic} (non-scattered);
 \item unstructured strong random background.
 \end{enumerate}
 The idea is very simple: backprojecting the incoming trajectories (or, in the Compton case, the whole surface cones of possible trajectories) of particles, one hopes that maybe, due to sufficient presence of ballistic particles detected from the source, one can see a statistically significant accumulation at the geometrically small source's location (see Fig. \ref{F:BP})

 Analysis done in \cite{ADHKK} provided a crude formula for the total number $N$ of particles (and thus observation time) needed to make detection with high (on the order of $99\%$) sensitivity and specificity (i.e., with low levels of false negatives and false positives).
  \begin{equation}\label{E:bpformula}
  N\gtrsim \left(\frac{8}{S}\right)p(1-p).
 \end{equation}
 Here $p$ is the ratio of the linear dimension of the source relative to the dimension of the cargo and $S$ is the SNR, defined as the proportion of the ballistic particles from the source versus the total number of source and background particles. In the cases considered in \cite{ADHKK} $N$ had to be on the order of $600000$, which is not unrealistic for $\gamma$ photons \textbf{not} screened by heavily shielding cargo.
 High specificity has been hardwired into the method, so satisfying (\ref{E:bpformula}) was only needed in \cite{ADHKK,cargo} to ensure high sensitivity.

 The implementation of the technique worked as follows \cite{ADHKK,cargo}: the data was backprojected, which resulted in a large background level throughout the volume. When the object was completely surrounded by detectors, this level was essentially constant and the mean was removed. When the detectors did not surround the object completely (e.g., no detector below the object), the global mean is irrelevant, and at each location the mean over a smaller patch was removed. After this clean-up the locations with an intensity less than five standard deviations above the mean suggested by the Central Limit Theorem were cut off. The results were interpreted as indications of a source being present. Thousands of Monte Carlo simulations showed that the inequality (\ref{E:bpformula}) agrees well with experiments and if $N$ is at or above this threshold, detection occurs with high sensitivity and specificity \footnote{An alternative Bayesian approach was implemented in \cite{X}.}.

\begin{figure}[ht!]
\centering
\begin{tabular}{c c c}
     \includegraphics[width=0.19\textwidth]{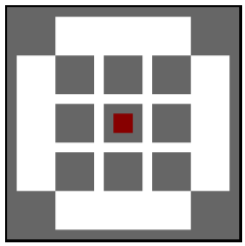} & \includegraphics[width=0.26\textwidth]{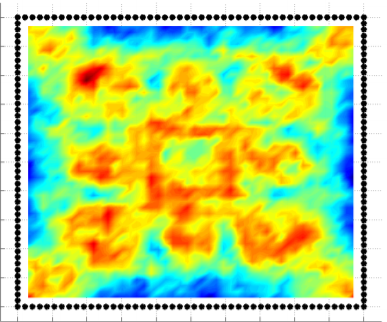} & \includegraphics[width=0.26\textwidth]{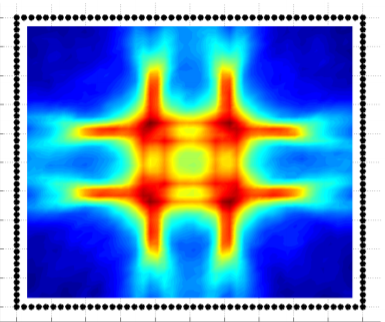}
\end{tabular}
\caption{(Left): Example of complex cargo configuration for which backprojection methods fail (i.e., no statistically suspicious locations are found). The red spot denotes the source location, the grey area represents iron and the white area represents air. (Middle): Backprojection results in absence of source. (Right): Backprojection results in presence of source.} \label{fig:Cargo}
\end{figure}

 This technique works reasonably well in the absence of complex cargo, but starts failing if such cargo is present \cite{cargo}, due to the second and third assumptions being inapplicable\footnote{This cargo problem is mostly non-existent when detecting neutrons coming from the source. However, some other (non-mathematical) issues arise, such as for instance lower number of particles detected.}. However, visual inspection of the backprojected data (see \cite{cargo}) seems to indicate that the data \textbf{might} still contain a signature of the source presence. Indeed, when the method of \cite{ADHKK} was applied to some cases of complex cargo in \cite{cargo}, despite its failure to detect presence of the source, such signatures (e.g., different highlighting of the pathways between cargo boxes) seemed to appear only when a source was present (see Figure \ref{fig:Cargo}). The reader should take into account that the color scales are different in the three pictures there and assigned automatically by the visualization software. This is of no importance, since it is not the intensity, but rather the patterns of highlighted pathways between boxes seem different.

No model of this effect has been developed, no telling features have been learned, and thus no detection algorithm came out of such observations.

This has led the authors to attempt deep learning for the source inference in the hope that a network could learn what we could not. Thus, our main goal is to detect the presence/absence of a source, not necessarily its location. If there is high probability of presence of the source, in practice one would check the cargo with other (hand-held) devices. However, one also needs to achieve high specificity, to avoid large numbers of false positives.

We describe now the structure of the article. Section \ref{S:compton} contains a brief description of the Compton type cameras and references to the known analytic approaches.
Success of deploying neural networks is predicated upon our access to sufficient data for neural network training. Thus, the first step - generating various complex cargo scenarios is described in Section \ref{S:cargo}. To avoid the inverse crime (overfitting), different processes of generating cargos are used for creating training and testing samples. Then, in absence of real data (which would require having weapons grade nuclear materials and physically creating thousands of different cargoes), we use (Section \ref{S:forward}) the technique of forward radiation transport simulations customarily used in nuclear engineering. As has been mentioned, the actual type of radiation is mathematically irrelevant, but to be close to real world scenarios and numerical parameter values, the case of $\gamma$-photons coming from an U-238 source and real world material parameters for cargo are used. The design of the network is described in section \ref{S:CNN}. The results are presented in Section \ref{S:results}. Additional remarks can be found in section \ref{S:remarks}. Acknowledgements are provided in section \ref{S:acknowledgements}. Some auxiliary tables, figures, and algorithm descriptions are located in the Appendix.


\section{Collimated and Compton $\gamma$-Cameras}\label{S:compton}

\begin{figure}[ht!]
\centering
\begin{tabular}{c c}
     \includegraphics[width=0.2\textwidth]{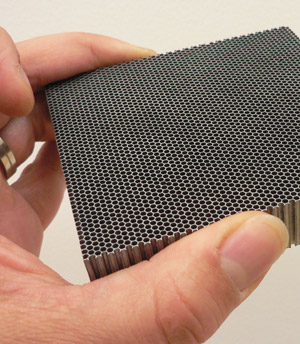} & \includegraphics[width=0.4\textwidth]{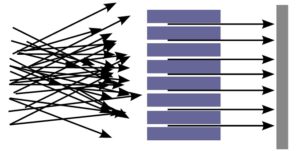}
\end{tabular}
\caption{(Left): Collimator used in nuclear medicine (Source: http://www.nuclearfields.com/collimators-nuclear-medicine.htm) (Right): Light collimation diagram (Source: http://www.fiber-optical-networking.com/getting-know-fiber-collimator.html)} \label{fig:Collimator}
\end{figure}

Mechanical collimators (see Figure \ref{fig:Collimator}) can be installed in front of a direction insensitive $\gamma-$camera to block all particles but those incident along (or close to) a desired trajectory.\\
Mechanical collimators are widely used in medical imaging. They, however, significantly attenuate the signal and require rotating the detector (or the object). In the applications with sufficiently high $SNR$, this additional data loss is not such a problem. In dealing with low $SNR$ signals however, this renders recovery of weak signals impossible. For this reason one can consider Compton type cameras instead.

\begin{figure}[h!]
\centering
\includegraphics[width=0.5\textwidth]{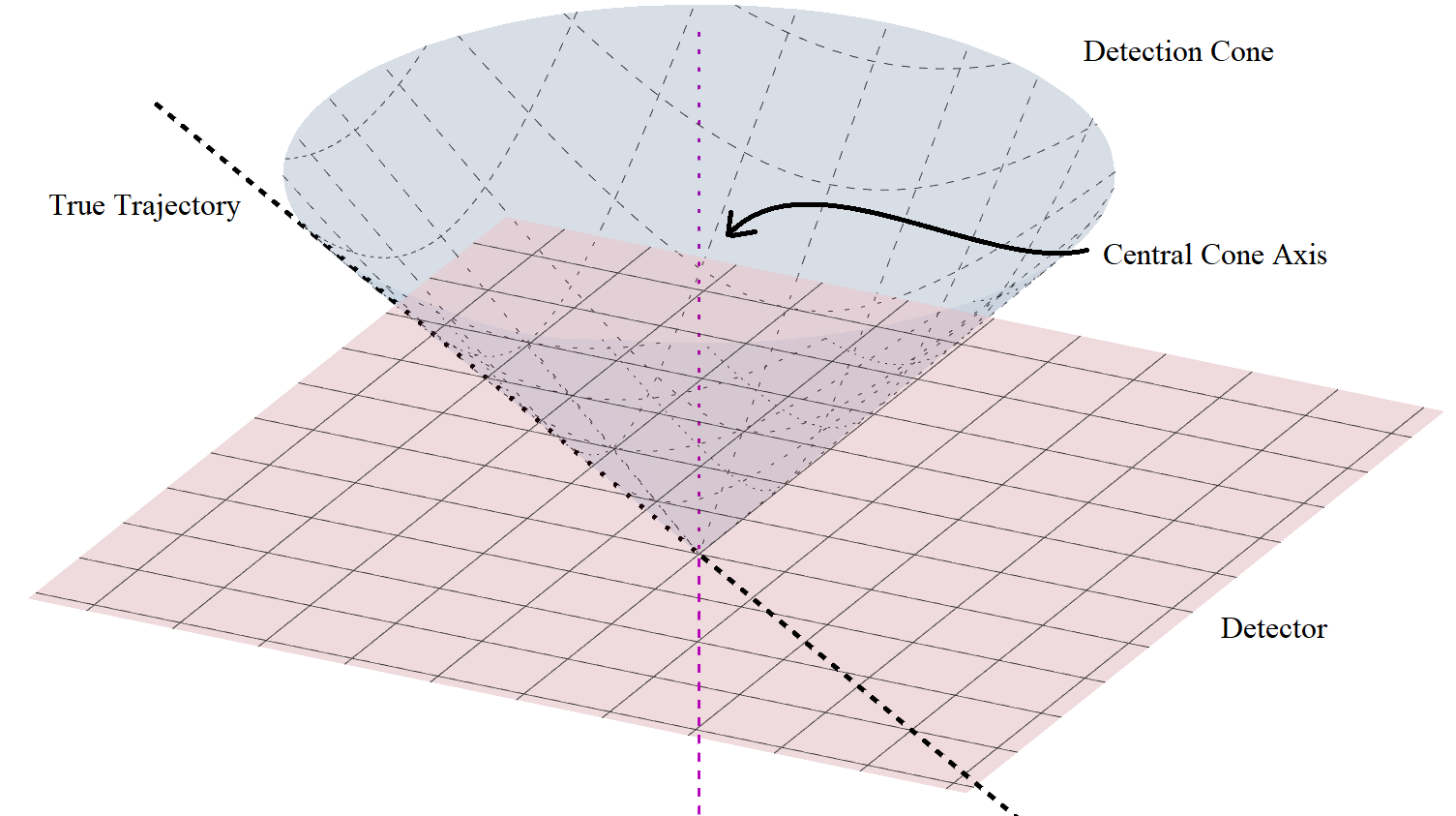}
\caption{Surface cone produced by Compton camera from particle detection} \label{fig:Compton}
\end{figure}

The \textbf{Compton camera }is a type of $\gamma$-particle detector\footnote{As we have mentioned before, novel neutron detectors (albeit based upon different physics rather than Compton scattering) that provide mostly similar cone information are currently being developed.} that does not attenuate the incident particles. The price to pay is that it provides less precise direction information than collimation would give. Namely, only a surface cone of possible incoming directions is measured rather than a precise trajectory (see Fig \ref{fig:Compton}).
In the absence of mechanical collimation, signal strength is effectively maintained, although the directional information is less precise and thus data analysis becomes more complex. On the other hand, the data provided is significantly over-determined (e.g., the space of cones in $3D$ is five-dimensional, versus the unknown distribution being three-dimensional). This turns out not to be a bad thing at all, but rather a blessing for stable inversion (see \cite{TKK}  for details and further references).

A variety of exact inversion formulas from Compton data of filtered-backprojection and other types have been developed and implemented (see \cite{TKK} and references therein). The choices are much more diverse than for the usual Radon transform inversions (see \cite{Natt}). The reason is that the Compton data is highly overdetermined. It was shown that this feature can be used to get high quality reconstructions in SPECT in presence of $50$\% noise and higher. However, this is a far cry from the low SNRs encountered in the homeland security problems described above.
\section{Simulating Cargo Scenarios}\label{S:cargo}

If one intends to tackle a problem using deep learning, it is natural to start by acquiring large amounts of training and testing data.

In order to obtain rich training data for a neural network, at least thousands (better hundreds of thousands or millions) of cargo configurations are needed. The more training data can be obtained, the better. Due to the sensitive nature of the materials involved in this work, we are unable to procure real-world data, so we resort to synthetic simulation. The high computation costs of these simulations restricted us to several thousands of samples, reaching up to $4\times 10^5$. However, our results (see Section \ref{S:results}) already show a success in detection.

To start, we randomly produce several thousand cargo configurations and compute forward radiation data simulations with up to four randomly placed sources and without them for each one. In order to avoid overfitting (and an inverse crime), different cargo generation procedures are used for producing the training and testing data.

\subsection{Procedural Generation of Training Cargo Configurations}

 A square cargo hold of size of $2.4m \times 2.4m$ is assumed and partitioned into $2.4cm \times 2.4cm$ cells (the possible source would occupy one of them). Each cell can be indexed via a pair of row and column indices, $(i,j)$, with $1 \leq i,j \leq 100$ and is assigned a material identification number $ID_{i,j}$. These numbers correspond to a variety of  materials, including Air, concrete, highly enriched uranium, iron, cotton, wood, plastic, and fertilized (their detailed chemical content described in \cite{cargo}).

Real cargo typically consists of several boxes with small spaces in between. In order to emulate this, an algorithm is implemented to generate different cargo configurations. It consists of three main steps:
\begin{itemize}
    \item A network of several horizontal and vertical ``corridors'' between boxes with random widths and locations is generated. The number of corridors $c$ is selected randomly in a desired range $c_{min} \leq c \leq c_{max}$.
    \item The resulting configurations are unlikely to be symmetric, while real cargo might happen to be symmetric. To check whether symmetry plays any role in detectability\footnote{Disclosure: Our results show that symmetries do not influence detectability.}, a portion of the samples produced are ``symmetrized'' by enforcing various (rotation and mirror) symmetry rules.
    \item Connected components of the rest of the space are identified as distinct ``cargo boxes.'' Then material contents are assigned to all boxes. In a subset of (rather than all) symmetric cargo configurations, material contents are also ``symmetrized'' according to the corresponding rule.
\end{itemize}

Generating the corridors between pieces of cargo is performed using a modification of the procedure outlined in \cite{roads} for generating road networks. For the training set, we only use networks consisting of horizontal and vertical segments, while for the testing set tilted and non-orthogonal pathways are allowed.
\begin{remark}
Instead of selecting corridor locations uniformly randomly, their locations for training are selected according to a probability distribution generated from a type of gradient noise developed in \cite{Perlin} in order to automate the production of realistic looking textures in computer graphics. A different algorithm is used for testing samples.
\end{remark}
Identification of connected components (``boxes'') is performed using SciPy's (Scientific Python, a popular Python package for scientific computing \cite{scipy}) implementation of the algorithms outlined in \cite{CC}.

The entire generation procedure is summarized in Algorithm 1 in the Appendix (Section \ref{S:Appendix}).

\begin{figure}
    \centering

    \begin{tabular}{ccccc}
        \includegraphics[width=2.75cm]{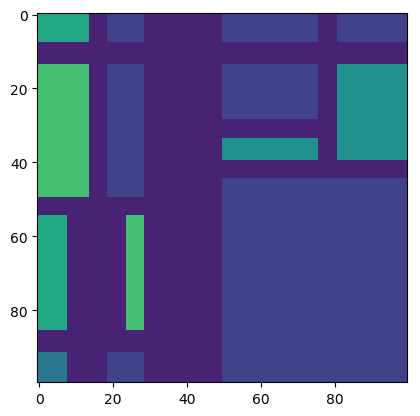} & \includegraphics[width=2.75cm]{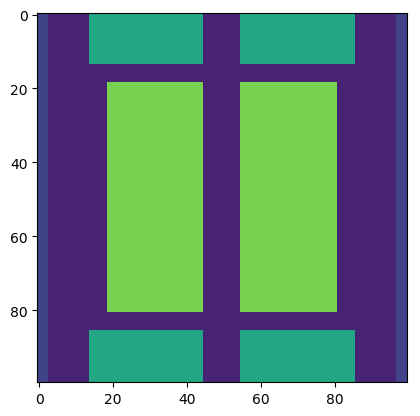} & \includegraphics[width=2.75cm]{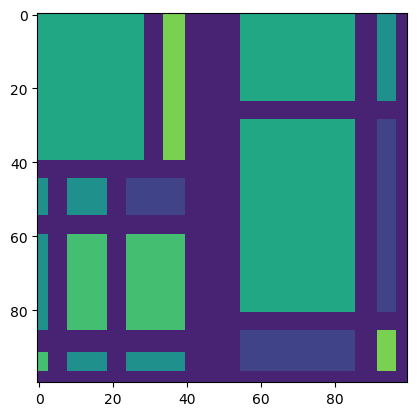} & \includegraphics[width=2.75cm]{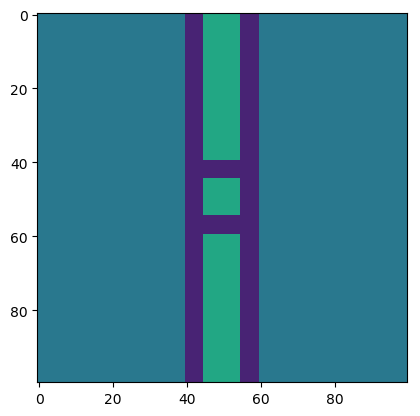} & \includegraphics[width=2.75cm]{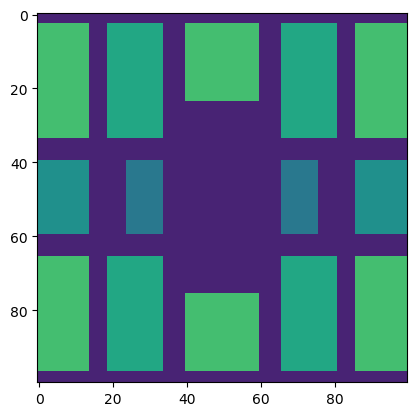}  \\
        \includegraphics[width=2.75cm]{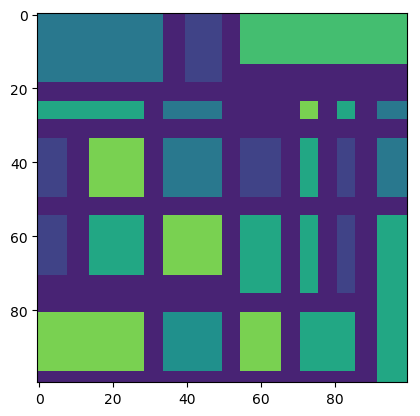} & \includegraphics[width=2.75cm]{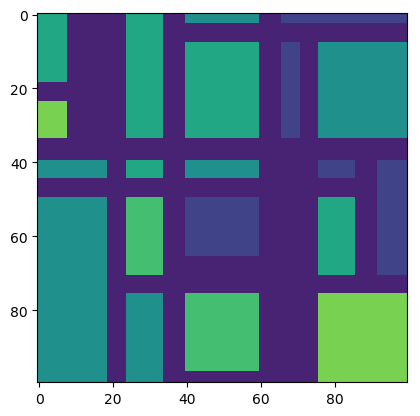} & \includegraphics[width=2.75cm]{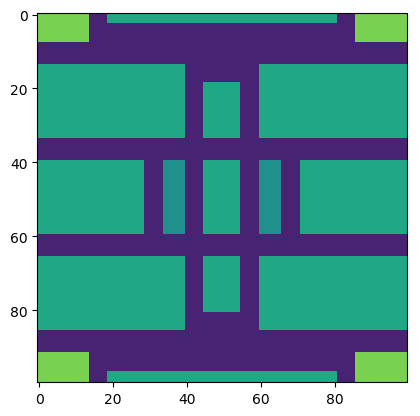} & \includegraphics[width=2.75cm]{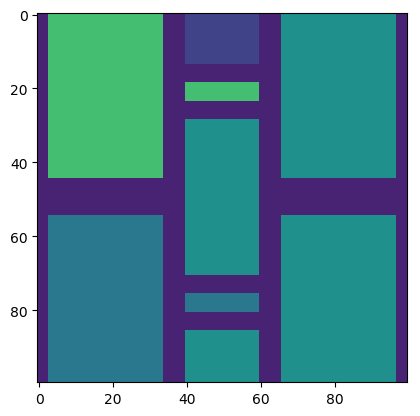} & \includegraphics[width=2.75cm]{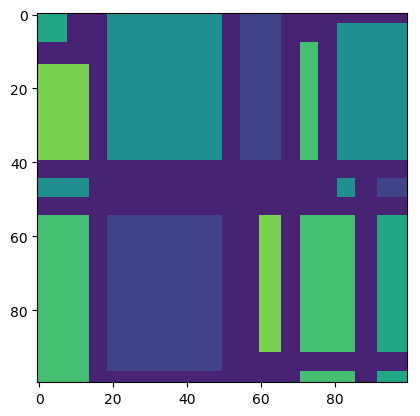}
    \end{tabular}
    \caption{A selection of cargo configurations procedurally generated via Algorithm 1}
    \label{fig: examples}
\end{figure}

\subsection{Procedural Generation of Testing Cargo Configurations}
To avoid the inverse crime of overfitting, testing configurations are produced by a somewhat similar, but independent algorithm. Namely, the middle points, the lengths and width of the corridors are selected randomly and independently. Moreover, the corridors are not required to be vertical or horizontal, or even orthogonal at their intersections anymore. Finding the boxes (connected components of the complement) and filling them with materials is also done randomly, similarly to the training case.

\subsection{Source placement}
A source of a (randomized) strength corresponding to approximately $1\%$ SNR is placed randomly into the cargo.

Multiple sources ($0, 1, 2, 3$, or $4$) are also modeled to see the effect on detection. Two scenarios are used:
\begin{enumerate}
\item when all the sources have the same strength $\approx 1\%$ SNR\\
and
\item when the strength of the source is diluted between several locations.
\end{enumerate}
One naturally expects deterioration of the detection in the 2nd case, while \emph{a priori} it would not be surprising if it happened in the 1st as well (although our results will demonstrate that this does not happen). Indeed, the backprojection detection, as well most probably the one by deep networks, if successful, should use some geometric assumptions (e.g., geometric smallness of the source), since the source's strength alone would not be statistically significant. Thus, multiplying the number of sources in principle might degrade the geometric features of importance (albeit one does not know what these are).

\section{Forward radiation simulations}\label{S:forward}
As previously mentioned, the nature of particles is irrelevant, but in order to be in realistic situations, the $\gamma$ particle detection is considered, where the material parameters and emission and background rates that are used assume realistic values.

After the cargo scenario has been created, one needs to simulate training and testing data by solving a massive forward radiation transport computation. Luckily, this is what nuclear engineering researchers are trained to do.

\subsection{Physics Preliminary}\label{S:physics}
U-238 (Uranium-238) photons from the 1.001 MeV emission line have mean-free-path in high-Z materials sufficiently high to be detected outside the container
(13.3mm mean-free-paths) \cite{Santi2013PassiveNA}.
In our application, sources of background radiation include a concrete base located some distance below the container. (Cosmic rays and other natural sources can be easily included and do not influence the results much.) These background sources radiate at much higher energies than 1.001 MeV, including 1.461 MeV from Potassium-40, 1.12 MeV and 1.76 MeV from Bismuth-214, and 2.61 MeV from Thallium-208 (Bismuth and Thallium are products of the decay of Uranium-238 and Thorium 232 respectively, and are present in trace amounts in concrete). Gamma photons which downscatter from these sources into the energy group surrounding the 1.001 MeV line account for the noise in our signal.
Gamma photons from the source will also undergo scattering and absorption within the volume of the container, which will reduce the number of ballistic source particles reaching the detectors placed around the container, thus weakening the signal.

\subsection{Mathematics of the forward radiation data simulation}

The radiation transport within the cargo container is modeled by the linear Boltzmann equation, given below using the multigroup approximation:
\begin{equation}
    \vec{\Omega} \cdot \vec{\nabla} + \Sigma_{t}^{g}( \vec{r}) \Psi^{g} (\vec{r}, \vec{\Omega}) = \sum_{g' = 1}^{G} \sum_{l=0}^{L} \Sigma_{s,l}^{g' \rightarrow g} ( \vec{r} ) \sum_{m=-l}^{l} \Phi_{l,m}^{g'} (\vec{r}) + Q^{g} (\vec{r}, \vec{\Omega})
    \label{eq: Boltzmann}
\end{equation}
where $\vec{r} \in \mathcal{D}$ is the position, $\vec{\Omega} \in \mathbb{S}^2$ the set of discrete directions and $g \in [1,G]$ the energy group. $\mathcal{D}$ is the volume of the cargo container, $\mathbb{S}^2$ is the unit sphere, $G$ is the total number of energy groups, $\Psi^g$ is the photon angular flux in the energy group $g$, $\Sigma_{t}^{g}$ is the total interaction cross section in group $g$, $\Sigma_{s,l}^{g' \rightarrow g}$ is the $l^{th}-$Legendre moment of the scattering cross section from group $g'$ to group $g$, $L$ is the maximum anisotropy expansion order, and $Q^g$ is the volumetric source of photons in group $g$ (stemming from the U-238 source). The moments of the angular flux are given by
\begin{equation}
    \Phi_{l,m}^{g}(\vec{r}) = \int_{4 \pi}  Y_{l,m} ( \vec{\Omega}) \Psi^{g}(\vec{r},\vec{\Omega})d \Omega
    \label{eq: moments}
\end{equation}
where $Y_{l,m}$ is the spherical harmonic of order of $l$ and degree $m$. Eq. (\ref{eq: Boltzmann}) is supplied with boundary conditions:
\begin{equation}
    \Psi^{g} ( \vec{r}, \vec{\Omega}) = h^{g}( \vec{r}, \vec{\Omega}) \qquad \forall \vec{r} \in \partial \mathcal{D}^{-}
    \label{eq: bd_cond}
\end{equation}
where $\partial \mathcal{D}^{-}$ is the incoming boundary defined as $\partial \mathcal{D}^{-}=\{ \vec{r} \in \partial \mathcal{D} \text{ such that } \vec{\Omega} \cdot \vec{n} ( \vec{r} ) < 0 \}$ with $\vec{n} ( \vec{r} )$ the outward unit normal vector at position $\vec{r}$. The function $h^g$ describes the background radiation due to a large concrete slab underneath the container, as previously described. Cross sections for various materials were generated using NJOY-99 \cite{MacFarlane1994TheNN}. The multigroup structure employed ranges from $1.00099$ MeV to $2.61449$ MeV with narrow bands centered at the radiation lines of the background and U-238.

For the purposes of this paper, calculations are carried out in two-dimensional space and only the energy group corresponding to the 1.001 MeV line is considered after solving Eq. (\ref{eq: Boltzmann}). The photon transport equation, Eq. (\ref{eq: Boltzmann}), is discretized using standard techniques:
\begin{enumerate}
    \item $S_n$ product Gauss-Legendre-Tchebychev angular quadrature \cite{Snchez2011OnTC} is employed (only a small number of polar angles are needed, but a very high number of azimuthal angles are needed to resolve properly the angular distribution in the 2D domain.)
    \item Spatial discretization based on a standard bilinear discontinuous finite element technique with upwinding at cell interfaces. \cite{reed1973triangularmesh,Wareing}
    \item Transport sweeps and Source Iteration are employed to solve the resulting system. \cite{Lewis}
\end{enumerate}
Once the transport equation (\ref{eq: Boltzmann}) has been solved, the outgoing angular photon flux at any boundary edge in 2D is recorded, which serves as the input data for use in Deep Learning and Backprojection.

Once configurations have been generated, a radiating source emitting an expected $8042.17$ photons per second at $1.001$ MeV is randomly placed, a forward radiative transfer equation is solved, and from its solution the radiation angular flux distribution on the boundary of the cargo is collected.

Due to linearity of (\ref{eq: Boltzmann}), the situations of presence of zero to four randomly placed sources could (and were) easily incorporated.

\section{Convolutional Neural Network}\label{S:CNN}
Using a fully connected network for the problem seems to be hardly feasible even in 2D, less so in 3D, in particular due to high dimensionality of the Compton camera data. The saving grace here is that, as in many imaging problems \cite{DLSurvey}, one expects that the important correlations occur mostly between close pixels, and hence convolutional neural networks, which are much more compact due to weight sharing, offer a hope. We thus construct, train, and test a deep convolutional neural network (CNN). This hand-waving argument for using CNN needs to be confirmed by computations, which is done in this text.

The suggested CNN architecture is summarized in Figure \ref{fig:NNarch} below. The input data dimension is $144\times 10^3=400 \times 360 \times 1$, as we model 400 equally spaced detectors with 360 equally spaced angular bins and only one energy bin is used. The network is trained on $1689$ unique simulated cargo configurations with varying numbers of sources present. By exploiting the fact that the Boltzmann equation (\ref{eq: Boltzmann}) is linear, we can produce multiple new samples from each configuration by taking varying combinations of sources and detectors. We simulate up to four sources per configuration, and four linear arrays of detectors along each edge of the cargo. This leads to a total of $1689 \times 15 \times 16 = 405360$ total samples. The various combinations are summarized in Table \ref{tab:combos} Below.
\begin{table}[h]
    \centering
    \begin{tabular}{|c|c|c|c|c|c|c|}
    \hline
     \multicolumn{1}{|p{18mm}|}{\centering Number of Sources} & \multicolumn{1}{|p{18mm}|}{\centering One Detector} & \multicolumn{1}{|p{18mm}|}{\centering Two Adjacent Detectors} & \multicolumn{1}{|p{18mm}|}{\centering Two Opposite Detectors} & \multicolumn{1}{|p{18mm}|}{\centering Three Detectors} & \multicolumn{1}{|p{18mm}|}{\centering Four Detectors} & \multicolumn{1}{|p{18mm}|}{\centering Total}\\ \hline
    0 & 6756 & 6756 & 3378 & 6756 & 1689 & 25335 \\ \hline
     1 & 27024 & 27024 & 13512 & 27024 & 6756 & 101340 \\ \hline
     2 & 40536 & 40536 & 20268 & 40536 & 10134 & 152010 \\ \hline
     3 & 27024 & 27024 & 13512 & 27024 & 6756 & 101340 \\ \hline
     4 & 6756 & 6756 & 3378 & 6756 & 1689 & 25335 \\ \hline
     Total & 108096 & 108096 & 54048 & 108096 & 27024 & 405360 \\ \hline
    \end{tabular}
    \caption{Number of training samples in each category}
    \label{tab:combos}
\end{table}
The output of the CNN is two probability measures: $\mathbb{P}_d$ on $\{0,1\}$ and $\mathbb{P}_n$ on $\{0,1,2,3,4\}$. A source is determined to be present if $\mathbb{P}(x=1)>0.5$, and absent otherwise. $\mathbb{P}_n$ predicts the number of sources present, which we set to $k=\text{argmax}_{0 \leq j \leq 4} \mathbb{P}_n(x=j)$. The loss function used for training is the binary cross-entropy loss:
\begin{equation}
    \mathcal{L}(y,\hat{y}) = -y \log \hat{y} - (1-y)\log (1 - \hat{y}),
\end{equation}
where $y$ is the network prediction and $\hat{y}$ is the target value (see \cite{BCEL}). The CNN was trained on simulations of a localized source in the presence of high background noise ($SNR=0.01$).
In all cases, early stopping is used to halt training before over-fitting. The various hyper-parameter values used in training are summarized in Table \ref{tab:hyper-params} below. The CNN is implemented using Keras with Tensorflow as its backend.
Keras is a high level API (Application Programming Interface) for interfacing with machine learning toolkits such as Tensorflow, Theano, and Microsoft Cognitive Toolkit. It helps streamline the construction and training of neural networks \cite{Keras}. Tensorflow is Google's machine learning toolkit and was chosen due to its scalability, wide range of features, and the wide range of documentation and tutorials available \cite{Tensorflow}. Any parameters not explicitly mentioned here were set to default values.
\begin{figure}[h!]
\centering
\includegraphics[width=1.0\linewidth]{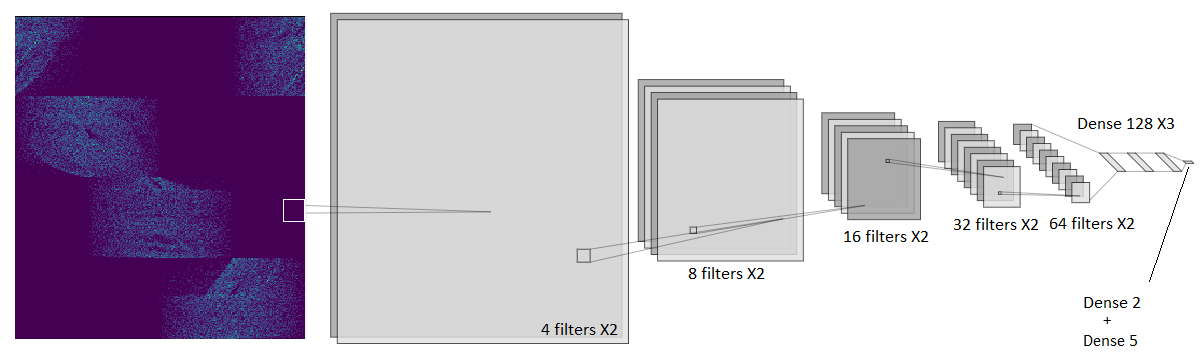}
\caption{CNN architecture used for source detection. The left-most cell shows an example of the detector data input to the CNN. $2 \times 2$ Max pooling layers are placed after every second convolutional layer.} \label{fig:NNarch}
\end{figure}
\begin{table}[h]
    \centering
    \begin{tabular}{|c|c|}
    \hline
    Optimization Method     &  Adam (See \cite{ADAM}) \\ \hline
    Activation     &  RELU (Softmax at output) \\ \hline
    Bias & True \\ \hline
    Convolution Window Size & 3x3 \\ \hline
    Learning Rate     & $2.0 \times 10^{-5}$ \\ \hline
    Learning Rate Decay Rate     & 0 \\ \hline
     Batch Size     & 4 \\ \hline
     Early Stopping Patience     & 3 epochs \\ \hline
     Loss     & Binary Cross-Entropy \\ \hline
    \end{tabular}
    \caption{Hyper-parameters used during training}
    \label{tab:hyper-params}
\end{table}

\section{Results}\label{S:results}

After training the CNN, we considered a large variety of cargo scenarios to test and to compare and contrast the performance of the CNN against the backprojection method of \cite{ADHKK,cargo}. We detail some interesting specific example scenarios in Sections \ref{S:Examples} and \ref{S:complex}. We then investigate the statistical performance of the CNN on large scale data sets to evaluate the sensitivity and specificity of the CNN in Section \ref{S: Statistical_Performance}, and to assess its performance with different numbers of sources and detectors in Section \ref{S:Number}. Finally, in Section \ref{S: Timing} we discuss the relation between cargo configuration and exposure time and how this affects the practicality of our technique.

\subsection{Example Scenarios} \label{S:Examples}

We describe now several (out of many, see later on in this text) sample results of testing the trained network on various scenarios not included in the training set.

\subsubsection {} \label{S:Ex1}
{\em Example \#1}

This configuration (right) as well as backprojection (left) with source present is shown in Figure \ref{fig:Ex1} below. The backprojection procedure described in the Introduction did not lead to any statistically significant detection. We, however, show the raw (not cleaned up) backprojection picture for the reader to notice the corridor highlighting phenomenon observed in \cite{cargo}. The network, on the other hand, succeeds in detecting presence of the source. This is one of the heavy iron configurations which has a shorter exposure time (18 seconds for 101,180 background particles).
\begin{figure}[ht!]
\centering
\begingroup
\addtolength{\tabcolsep}{-38pt}
\begin{tabular}{c c}
     \includegraphics[width=0.5\textwidth]{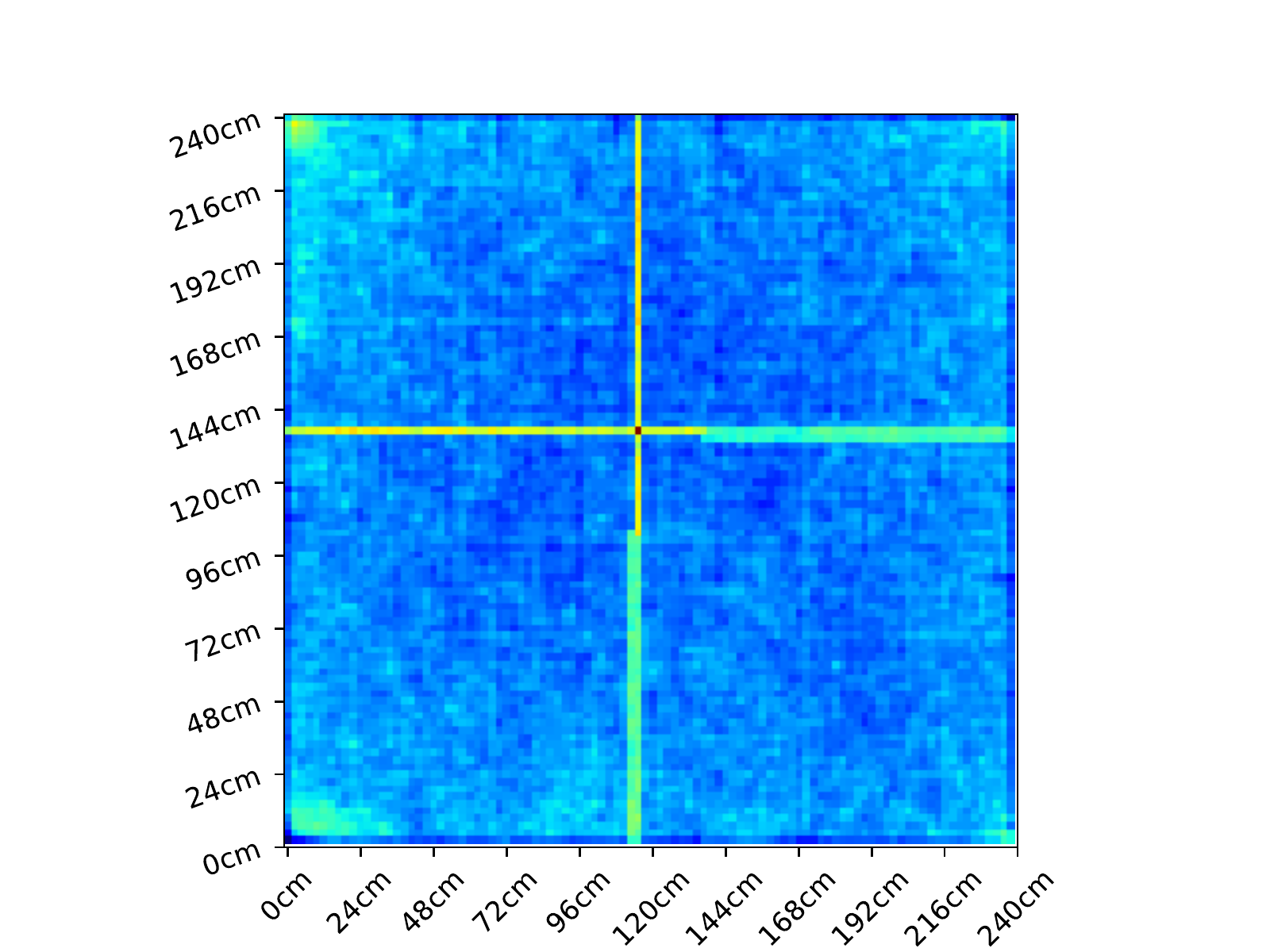} &
     \hspace{1.2cm}\includegraphics[width=0.5\textwidth]{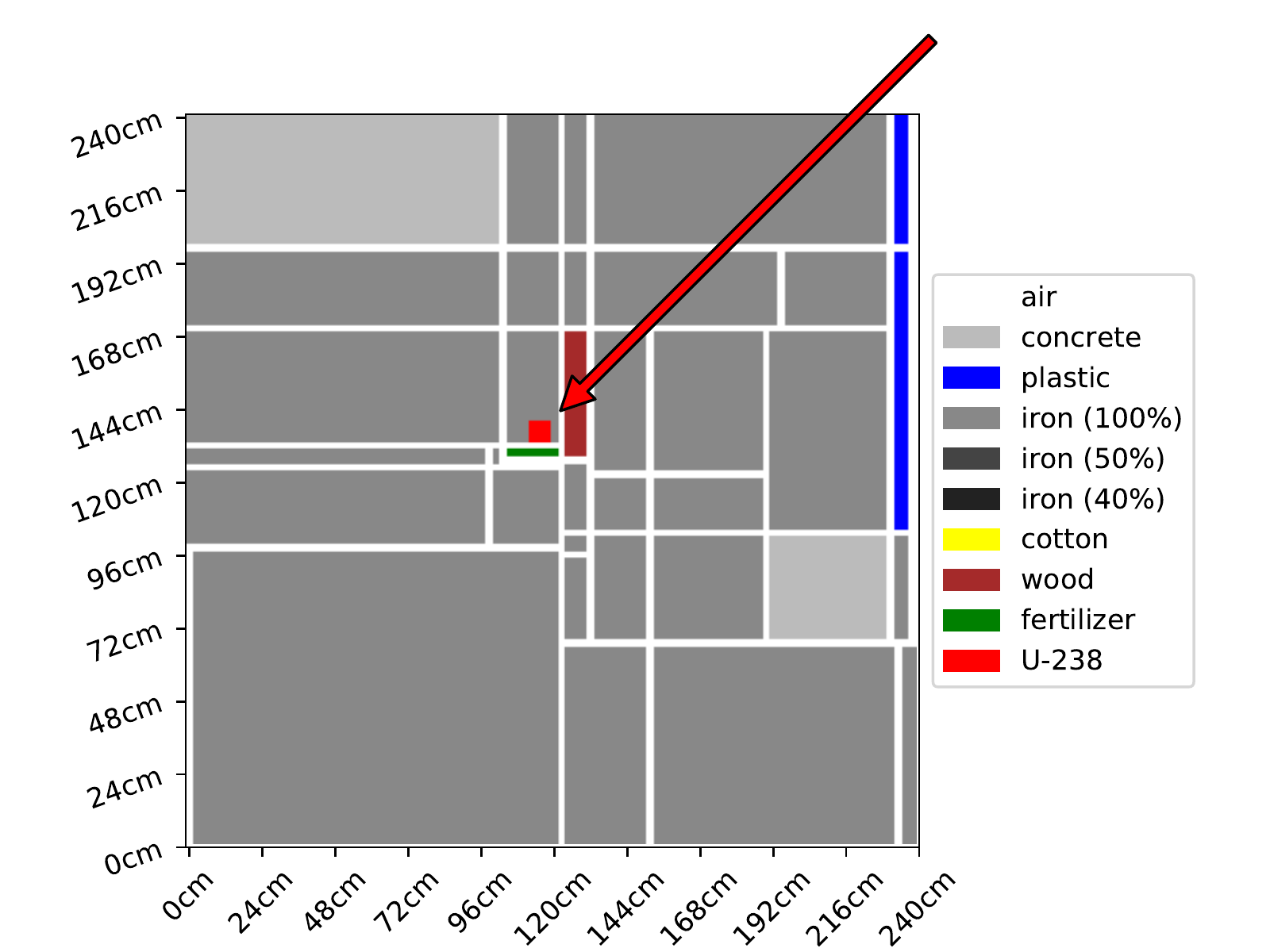}
\end{tabular}
\endgroup
\caption{Left: Backprojection with source detected. Right: Cargo configuration with source location indicated by arrow. 101,180 particles detected, 100,185 background particles and 995 source particles. Exposure time is 18 seconds.} \label{fig:Ex1}
\end{figure}

\subsubsection {} \label{S:Ex2}
{\em Example \#2}

\begin{figure}[ht!]
\centering
\begin{tabular}{c}
     \includegraphics[width=0.5\textwidth]{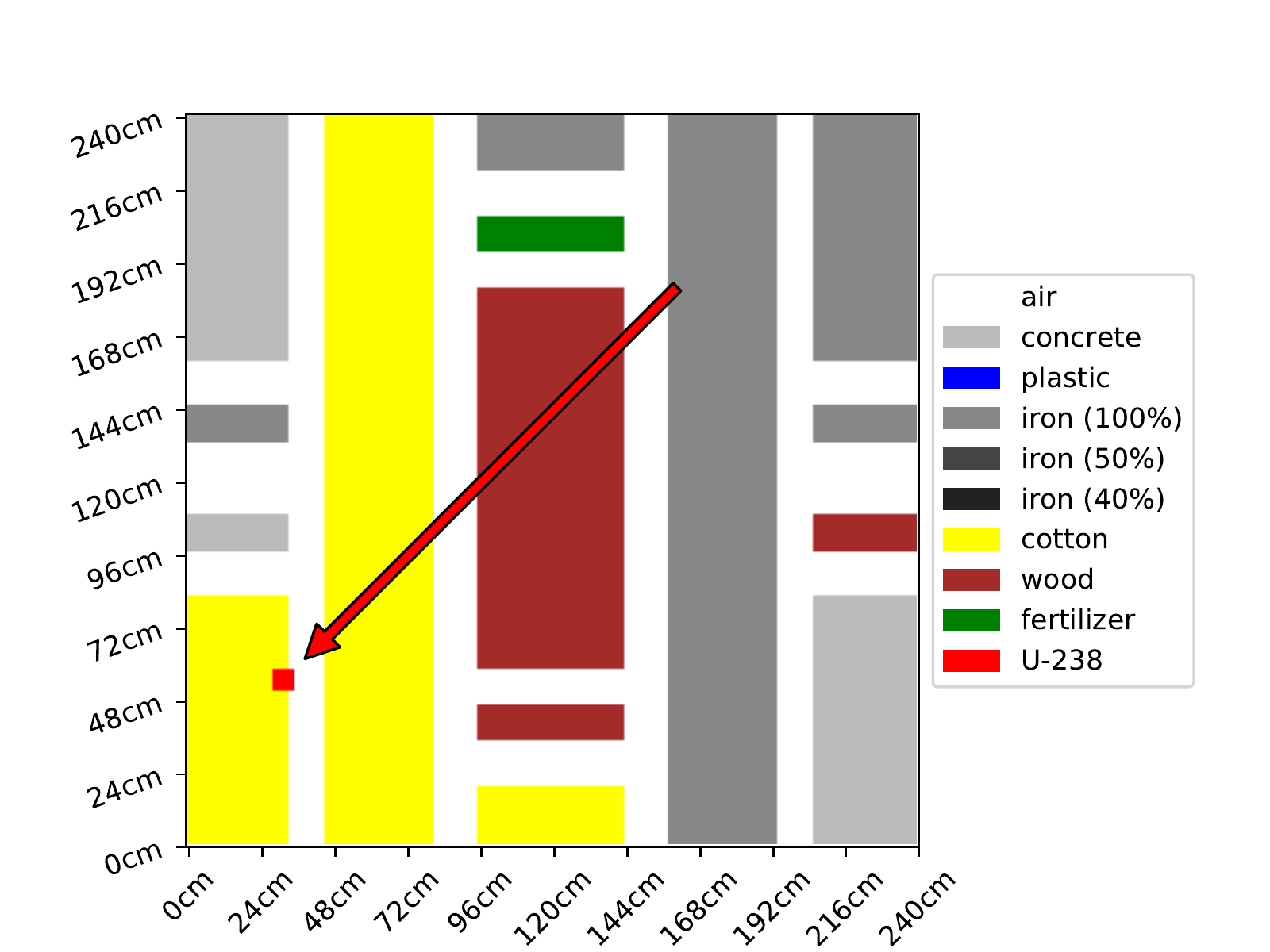}
\end{tabular}
\caption{Cargo configuration with source location indicated by arrow. 101,092 particles detected, 100,095 background particles and 997 source particles. Exposure time is 9 hours and 26 minutes.} \label{fig:Ex2}
\end{figure}

Next we consider the scenario shown in Figure \ref{fig:Ex2}, where backprojection fails to detect the source (and thus is not shown), but the network succeeds. Here the exposure time needed for the detection is significantly longer. In this configuration a long thick iron slab effectively blocks one side of the detectors. Smaller chunks of iron spread throughout the container further attenuate the signal along certain trajectories. As a result, it would take 9 hours and 26 minutes to detect the needed 101,092 particles. Unless one is talking about a shipping container, this is practically unfeasible. As the results in Section \ref{S: Statistical_Performance} show, twice shorter time would still do decently, and even five times shorter time might sometimes be used, although at the expense of higher false positive rate.

\subsubsection {} \label{S:Ex3}
{\em Example \#3}

\begin{figure}[ht!]
\centering
\begin{tabular}{c c}
     \includegraphics[width=0.5\textwidth]{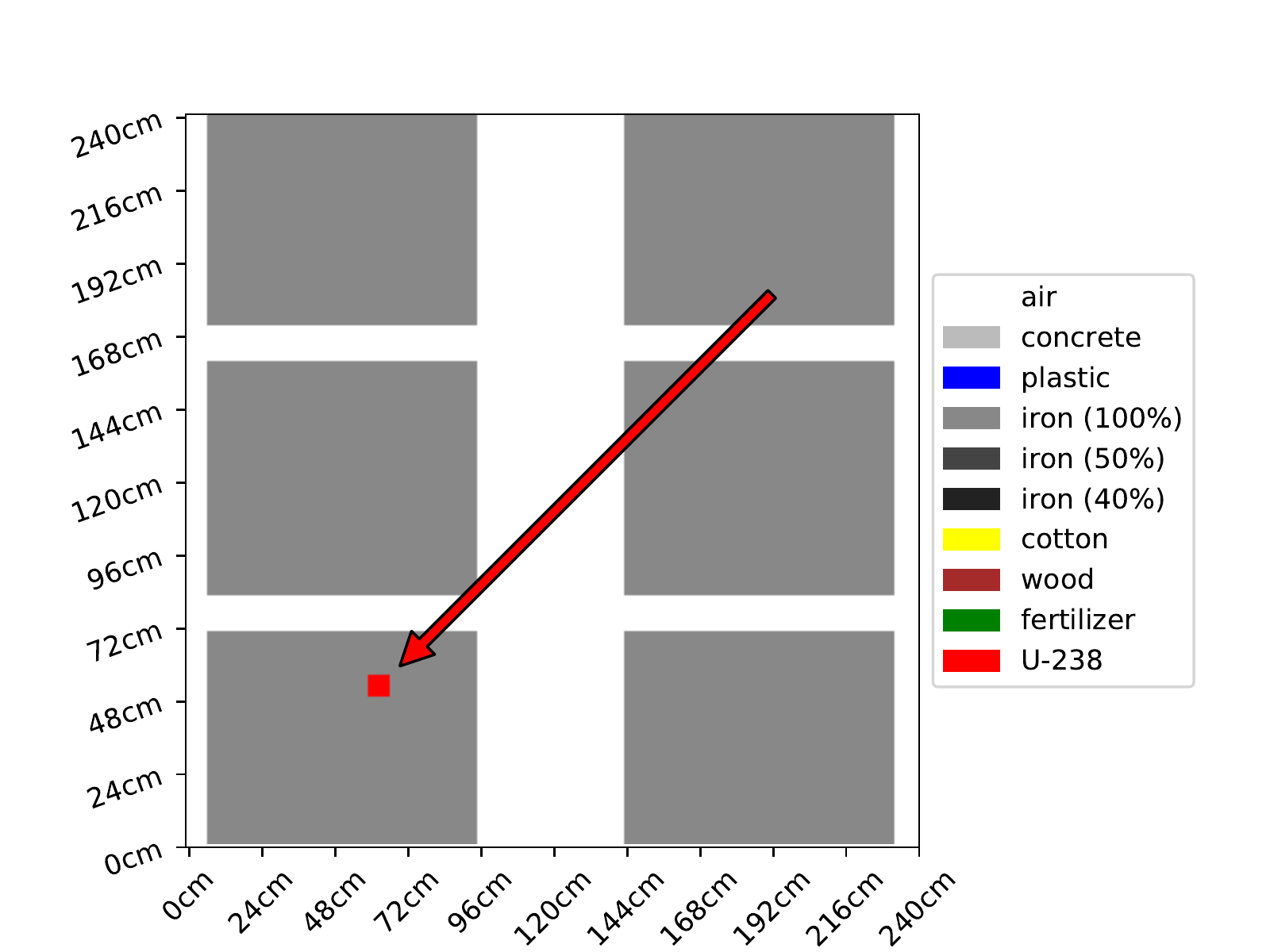}
\end{tabular}
\caption{Cargo configuration with source location indicated by arrow. 100,866 particles detected, 99,867 background particles and 999 source particles. Exposure time is 50 minutes and 17 seconds.} \label{fig:Ex3}
\end{figure}

Here we consider a somewhat more tenable scenario shown in Figure \ref{fig:Ex3}, where backprojection fails to detect the source, yet the network succeeds. In this case the exposure time is 50 minutes and 17 seconds for 100,866 particles. In this configuration several large blocks of iron are periodically tiled in the container, with the source located within one of the blocks.

\subsubsection {} \label{S:Ex4}
{\em Example \#4}

\begin{figure}[ht!]
\centering
\begingroup
\addtolength{\tabcolsep}{-38pt}
\begin{tabular}{c c}
     \includegraphics[width=0.5\textwidth]{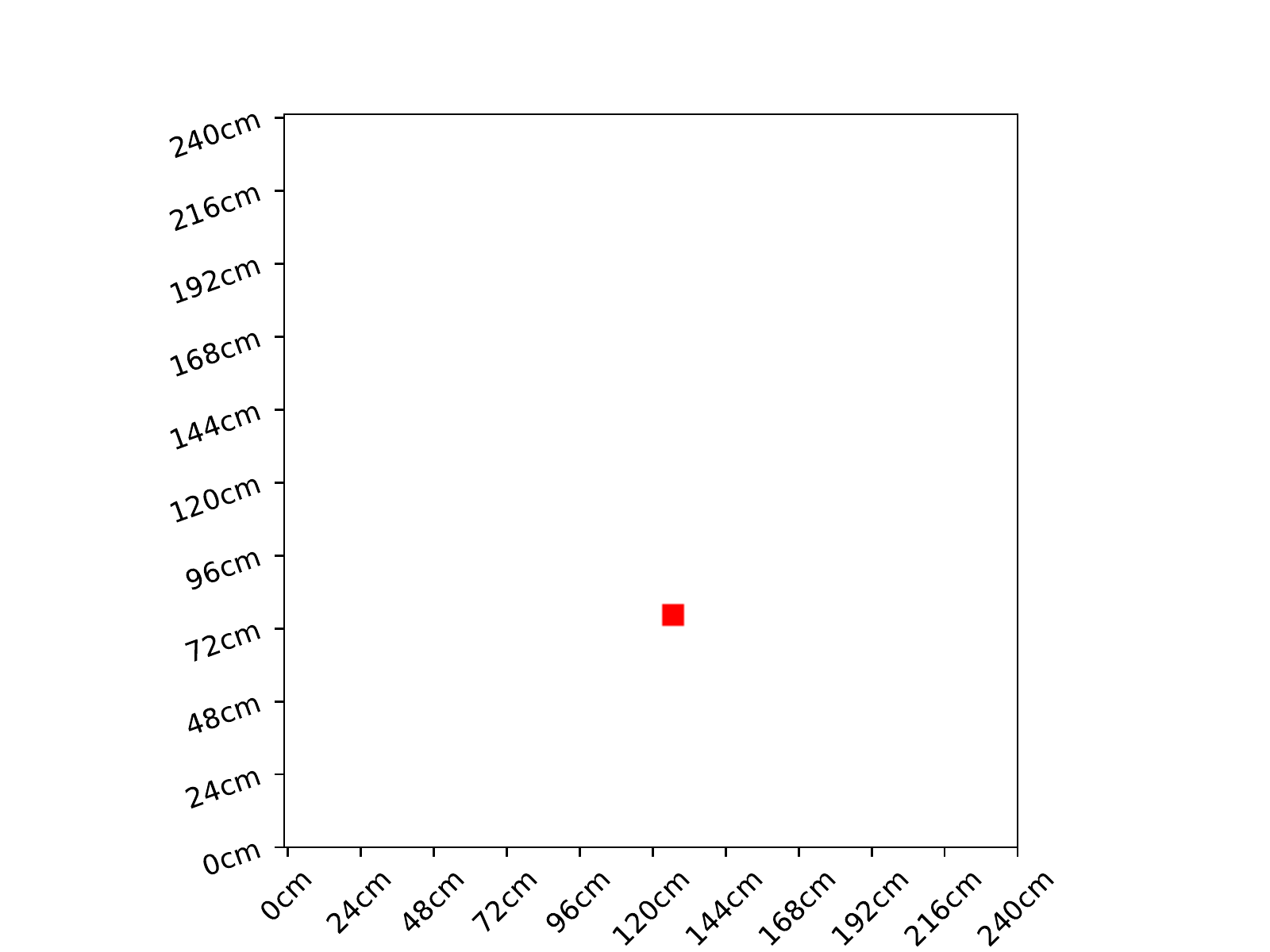} &
     \hspace{1.2cm}\includegraphics[width=0.5\textwidth]{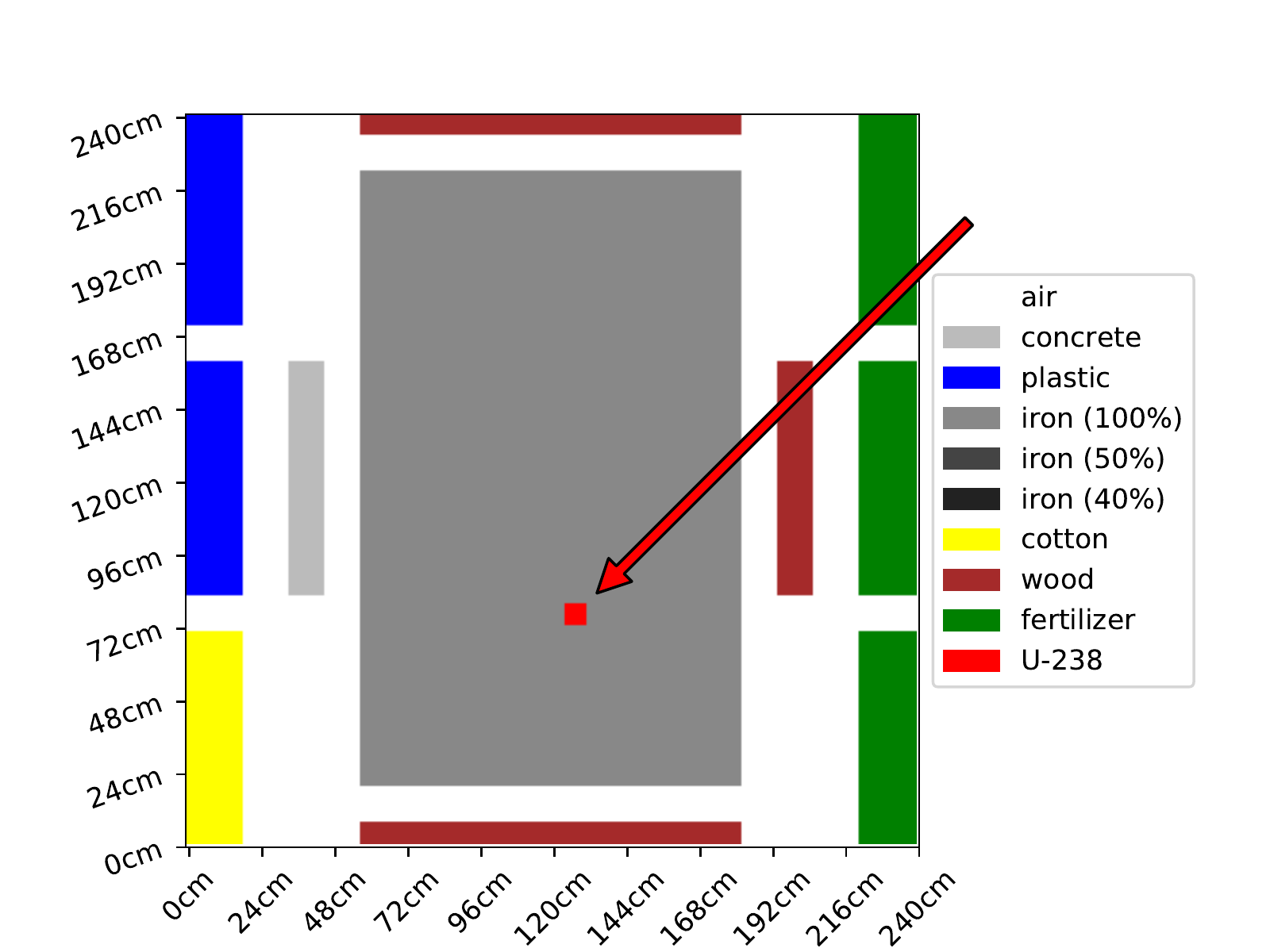}
\end{tabular}
\endgroup
\caption{Left: Backprojection with source detected. Right: Cargo configuration with source location indicated by arrow. 101,272 particles detected, 100,328 background particles and 944 source particles. Exposure time is 3 days and 12 hours.} \label{fig:Ex4}
\end{figure}

Now we consider a somewhat extreme scenario (Figure \ref{fig:Ex4}), where both approaches succeed in detecting the source. In this case the exposure time is 3 days and 12 hours for collecting 101,272 particles. In this configuration one very large block of iron in the center of the container surrounds the source. The source is still localized relatively well by backprojection for this scenario. Most of the cargo is filled with a homogeneous material, which might explain why backprojection did not fail.

\subsubsection {} \label{S:Ex5}
{\em Example \#5}

\begin{figure}[ht!]
\centering
\begingroup
\addtolength{\tabcolsep}{-38pt}
\begin{tabular}{c c}
     \includegraphics[width=0.5\textwidth]{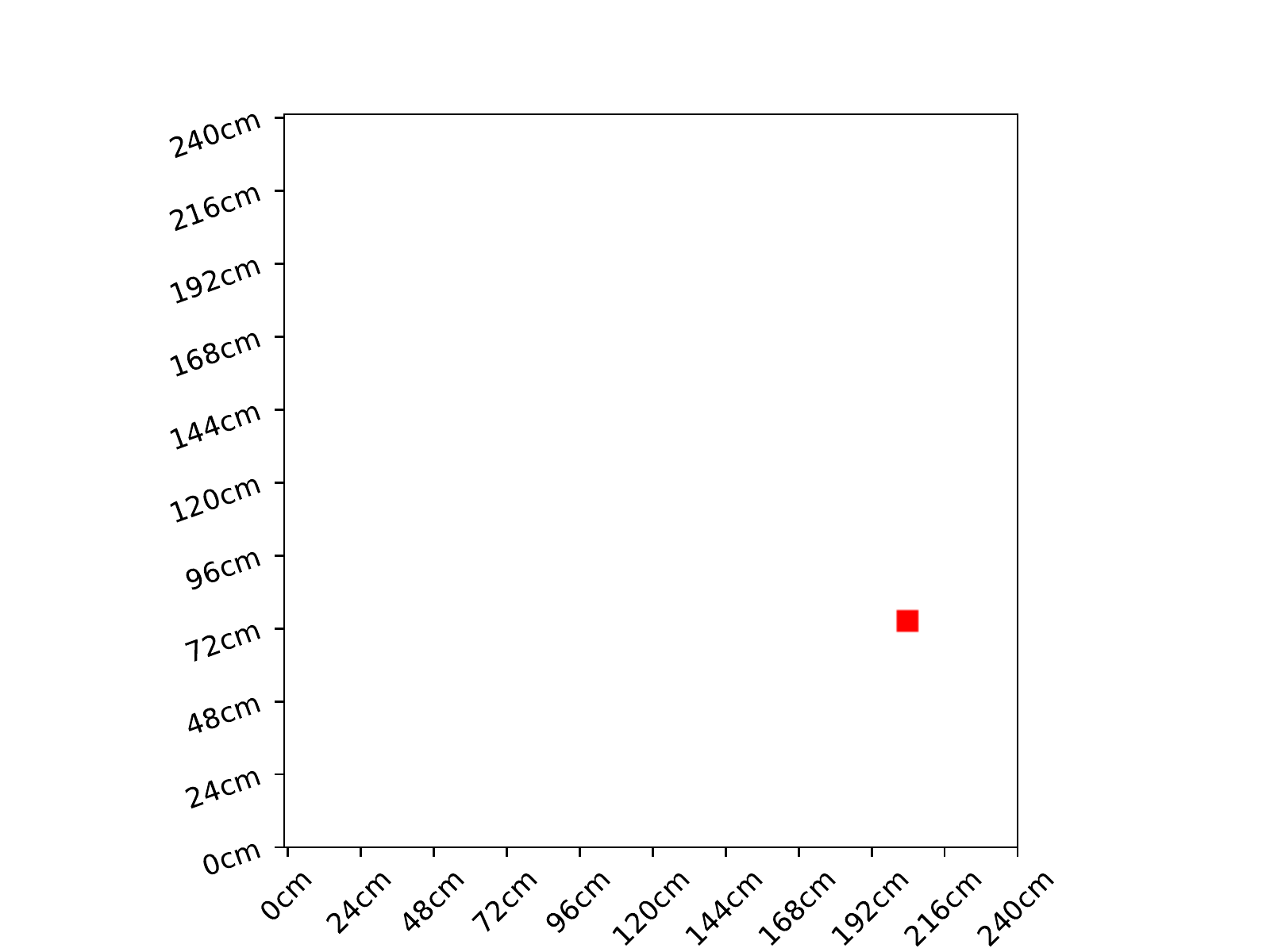} &
     \hspace{1.2cm}\includegraphics[width=0.5\textwidth]{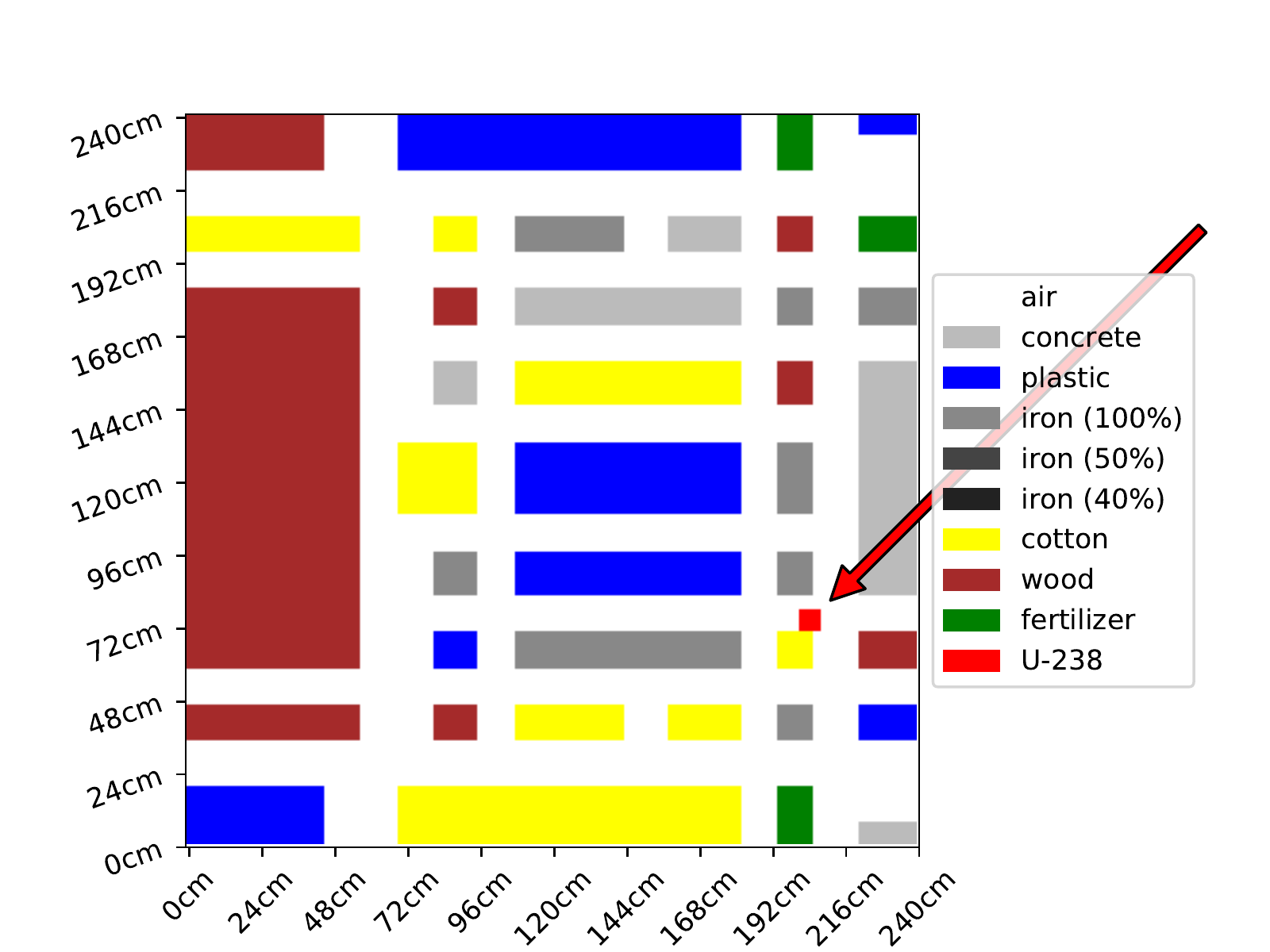}
\end{tabular}
\endgroup
\caption{Left: Backprojection with source detected. Right: Cargo configuration with source configuration indicated by arrow. 100,898 particles detected, 99,911 background particles and 987 source particles. Exposure time is 276milliseconds.} \label{fig:Ex5}
\end{figure}

Next, we consider a rather easy scenario (Figure \ref{fig:Ex5}), where both backprojection and the network succeed. In this case the exposure time is 276 milliseconds for 100,898 particles. In this configuration several small blocks of different materials are spread throughout the container. Only an insignificant amount of particles are scattered, so backprojection recovers the source distribution extremely well.

\subsection{Generalization to more complex scenarios}\label{S:complex}
Now we will include more complex situations, significantly different from the ones used for training. Namely, the corridors are not necessarily aligned vertically and horizontally, nor are intersecting corridors orthogonal. The algorithm of producing configurations was different from the one used in training. Additionally, we allow  multiple sources to be present. The results show that the network passes well this generalization test.

\subsubsection {} \label{S:Ex6}
{\em Example \#6}

\begin{figure}[ht!]
\centering
\begingroup
\addtolength{\tabcolsep}{-38pt}
\begin{tabular}{c c}
     \includegraphics[width=0.5\textwidth]{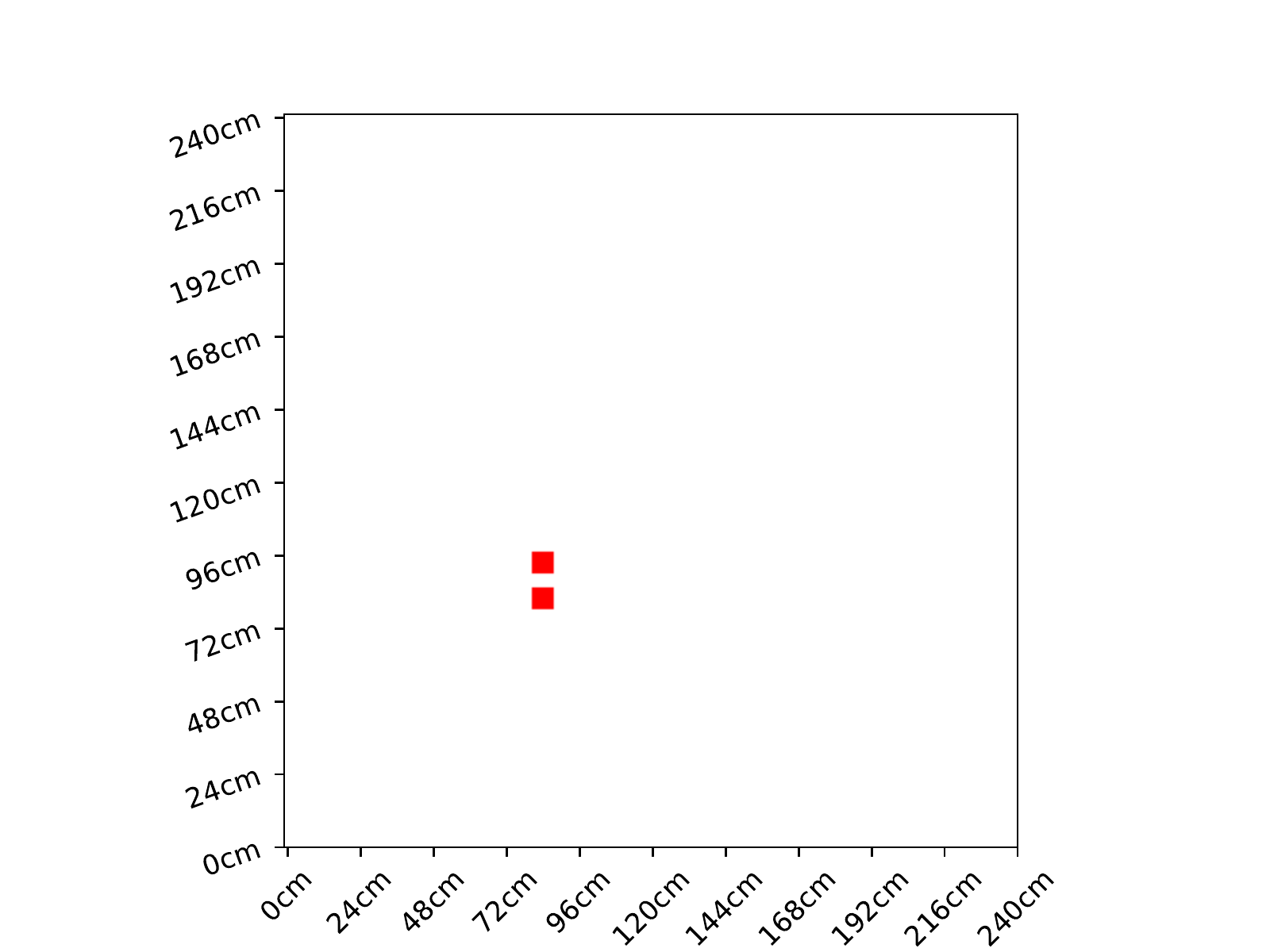} &
     \hspace{1.2cm}\includegraphics[width=0.5\textwidth]{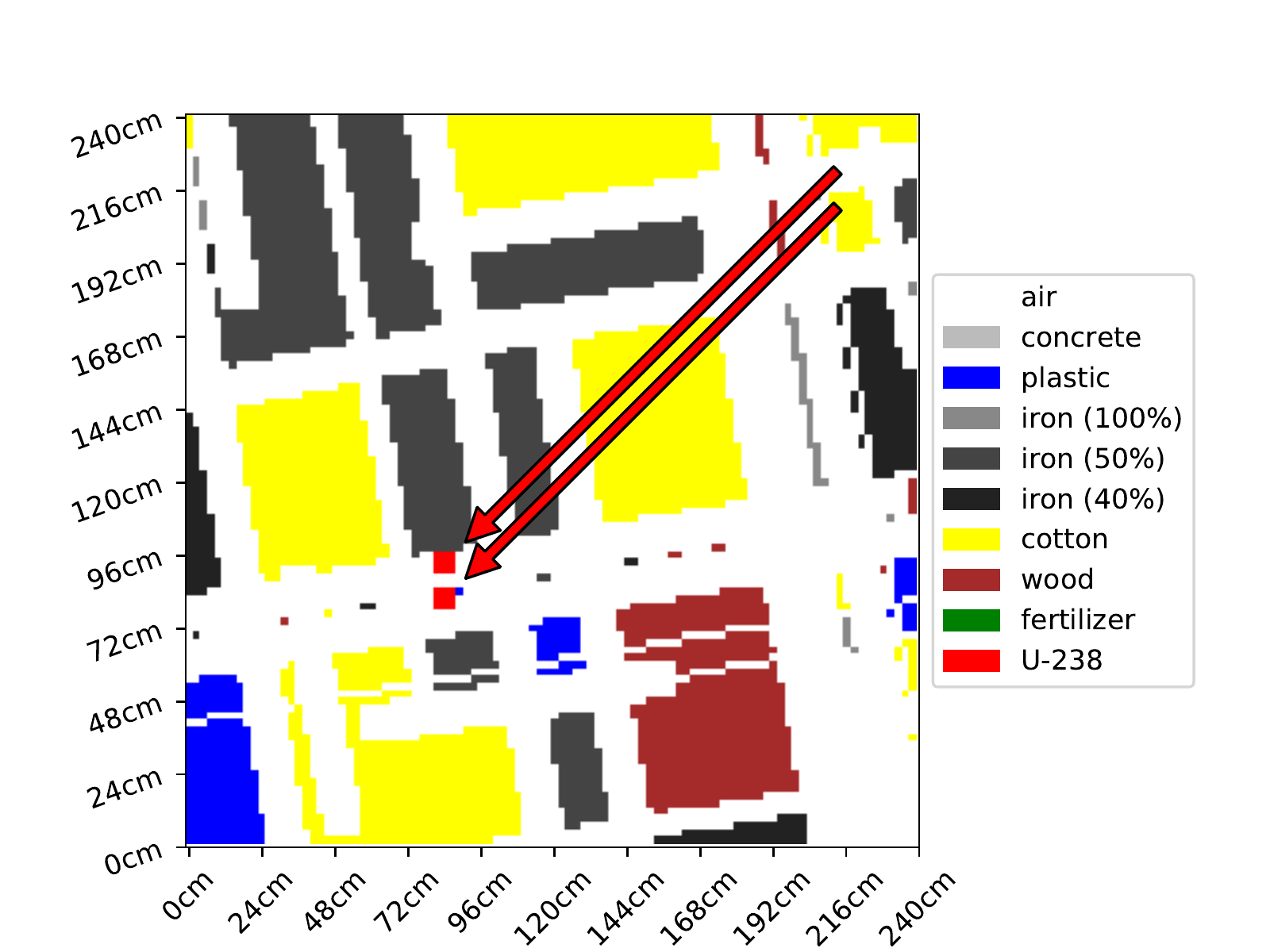}
\end{tabular}
\endgroup
\caption{Left: Backprojection with source detected. Right: Cargo configuration with source configuration indicated by arrow. 101,497 particles detected, 99,492 background particles and 2,005 source particles. Exposure time is 649 milliseconds.} \label{fig:Ex6}
\end{figure}

In this configuration (see Figure \ref{fig:Ex6}) several iron blocks are spread throughout the container, but a sufficient amount of low attenuating paths exist between the sources and detector arrays for backprojection to recover the sources well.
There are two sources present very near to each other. This clearly aids the backprojection method in successfully detecting the sources. The CNN also succeeds in detecting presence of both of the sources. Here the exposure time is 649 milliseconds for 101,497 particles.

\subsubsection {} \label{S:Ex7}
{\em Example \#7}

\begin{figure}[ht!]
\centering
\begingroup
\addtolength{\tabcolsep}{-38pt}
\begin{tabular}{c c}
     \includegraphics[width=0.5\textwidth]{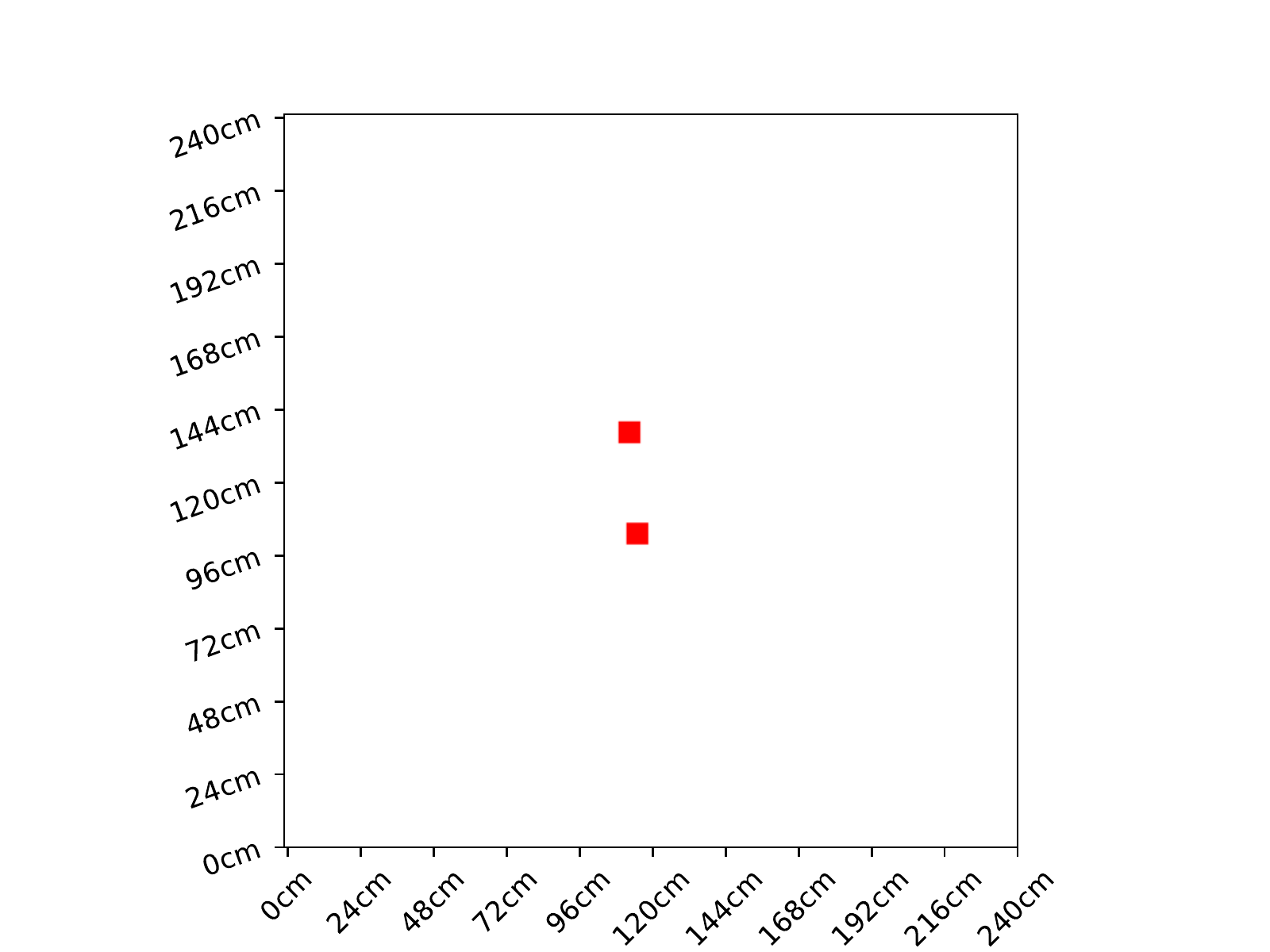} &
     \hspace{1.2cm}\includegraphics[width=0.5\textwidth]{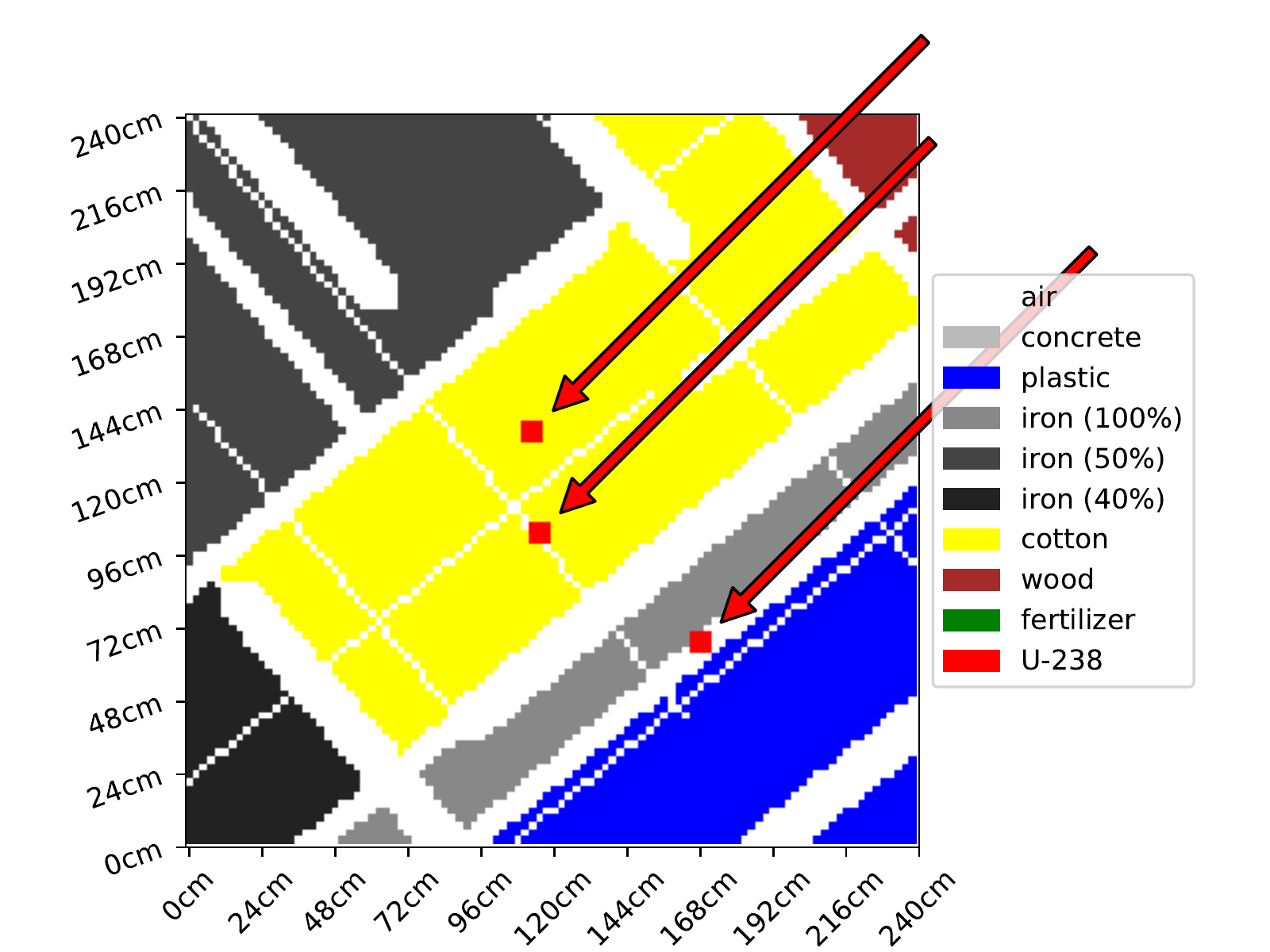}
\end{tabular}
\endgroup
\caption{Left: Backprojection with source detected. Right: Cargo configuration with source configuration indicated by arrow. 102,790 particles detected, 99,846 background particles and 2,944 source particles. Exposure time is 371 milliseconds.} \label{fig:Ex7}
\end{figure}

In this configuration (Figure \ref{fig:Ex7}) several heavy iron blocks cut diagonally through the container slightly off-center. Three sources are present in this scenario, the two sources around the middle are localized well with backprojection, since most of the materials only weakly attenuate the signal, but the source on the other side of the heavy iron has several attenuating materials to contend with, so the backprojection smears its signature throughout the diagonal corridor it's in. Both backprojection and the CNN successfully predict that there is a source, although backprojection fails to locate the third source. This third source may prove difficult for the CNN to contend with as well, as the CNN predicts there are only two sources present. Here the exposure time is 371 milliseconds for 102,790 particles.

\subsubsection {} \label{S:Ex8}
{\em Example \#8}

\begin{figure}[ht!]
\centering
\begingroup
\addtolength{\tabcolsep}{-38pt}
\begin{tabular}{c c}
     \hspace{1.2cm}\includegraphics[width=0.5\textwidth]{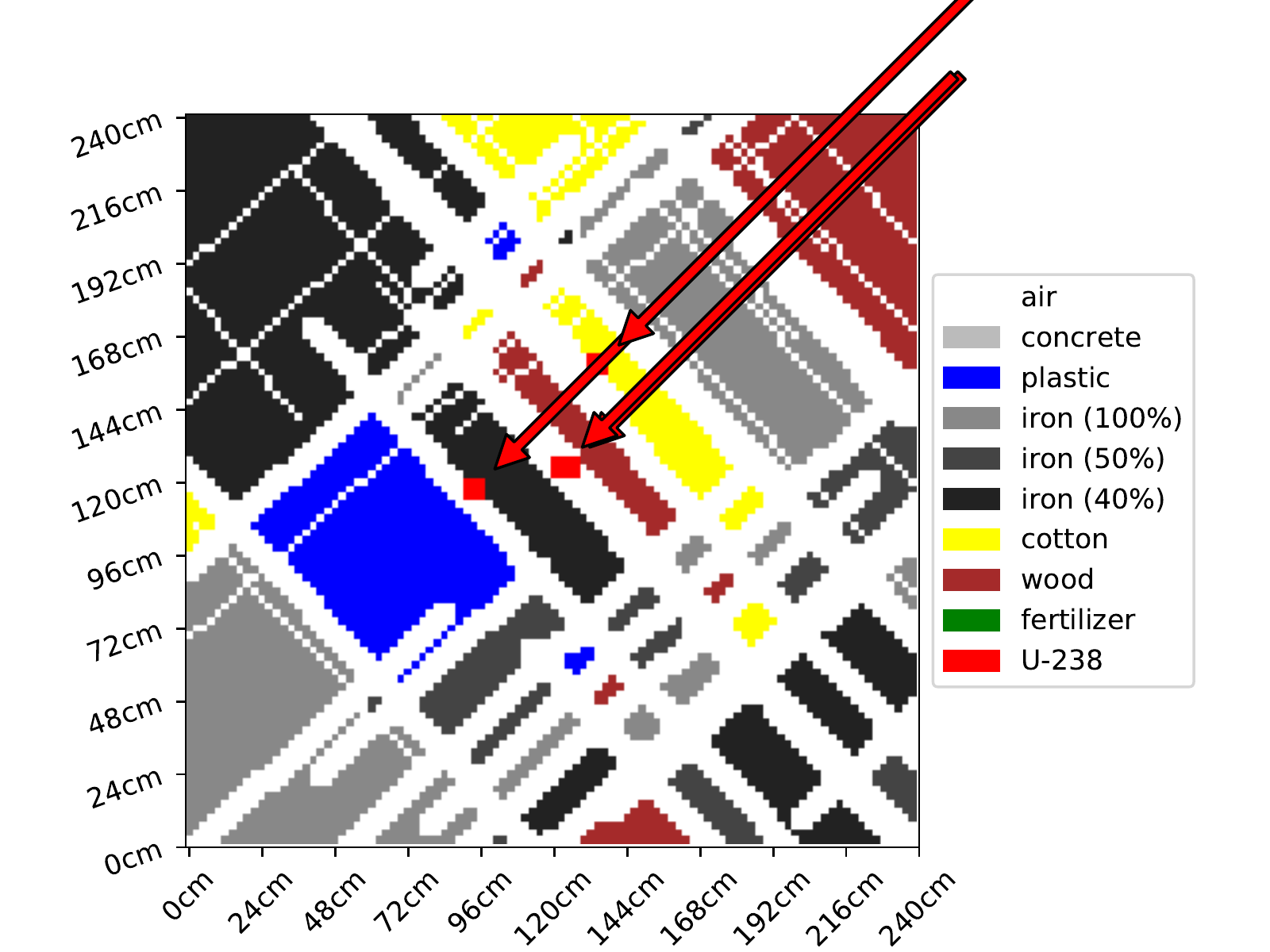}
\end{tabular}
\endgroup
\caption{ Cargo configuration with source configuration indicated by arrow. 103,789 particles detected, 99,900 background particles and 3,889 source particles. Exposure time is 1.97 seconds.} \label{fig:Ex8}
\end{figure}

Here several iron blocks surround the center of the container (Figure \ref{fig:Ex8}). Four sources are present in this scenario, two of them directly adjacent (and thus hard to distinguish in the picture) and all four are near the center of the container. In this case backprojection fails to localize any of the sources due the limited angular information in the signal as a result of the attenuating properties of the iron. The CNN, on the other hand succeeds in detecting the presence of all four of the sources, even despite the close proximity of two of them. Here the exposure time is 1.97 seconds for 103,789 particles.

\subsubsection {} \label{S:Ex9}
{\em Example \#9}

Finally, we consider a simple case where backprojection and the CNN both succeed. In this case there is ample angular information for backprojection to localize the source well and the CNN correctly predicts the presence of a single source. The exposure time is 21.92 seconds for 100,672 particles. The configuration can be seen in Figure \ref{fig:Ex9} below.
\begin{figure}[ht!]
\centering
\begingroup
\addtolength{\tabcolsep}{-38pt}
\begin{tabular}{c c}
     \includegraphics[width=0.5\textwidth]{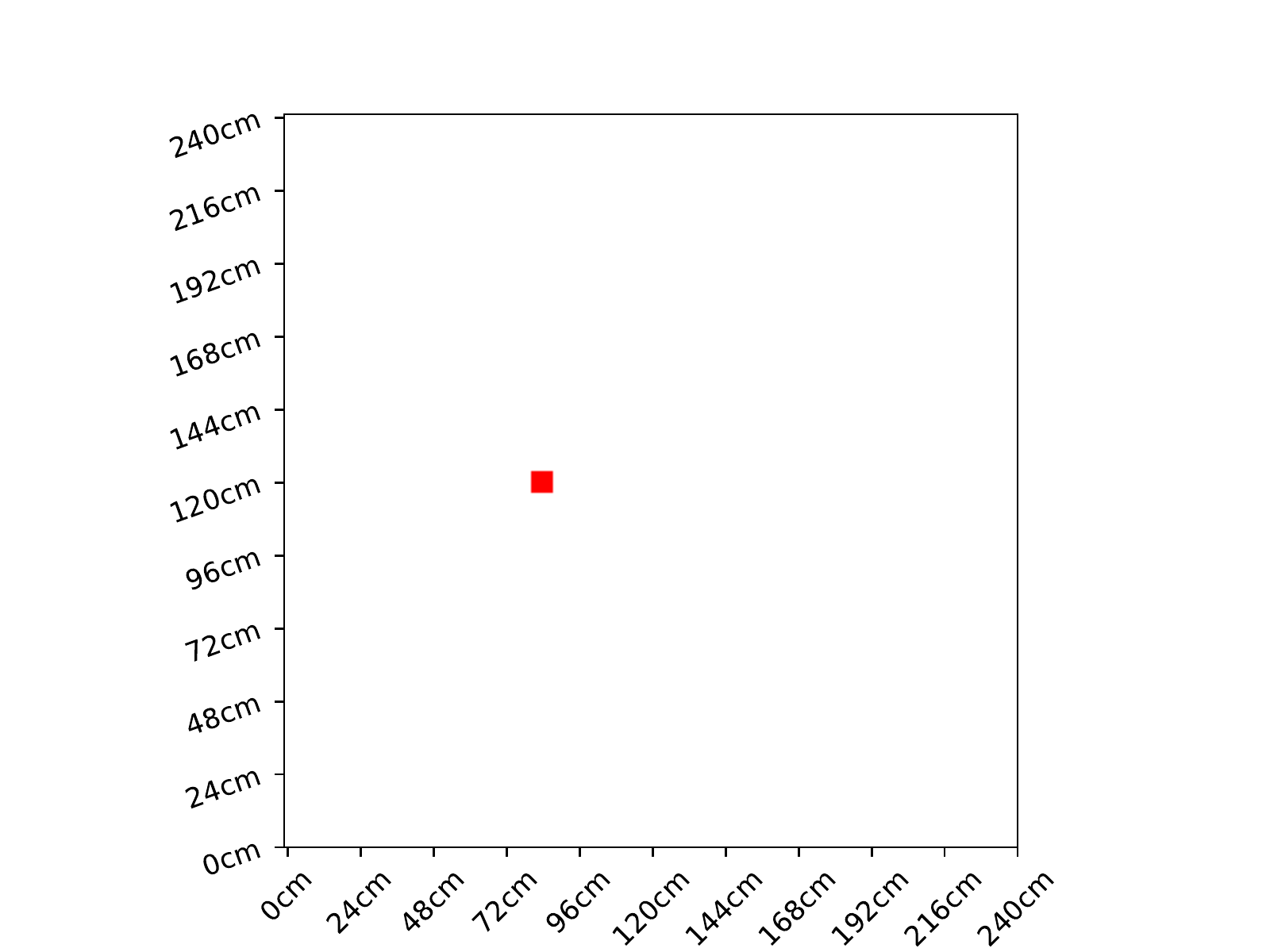} &
     \hspace{1.2cm}\includegraphics[width=0.5\textwidth]{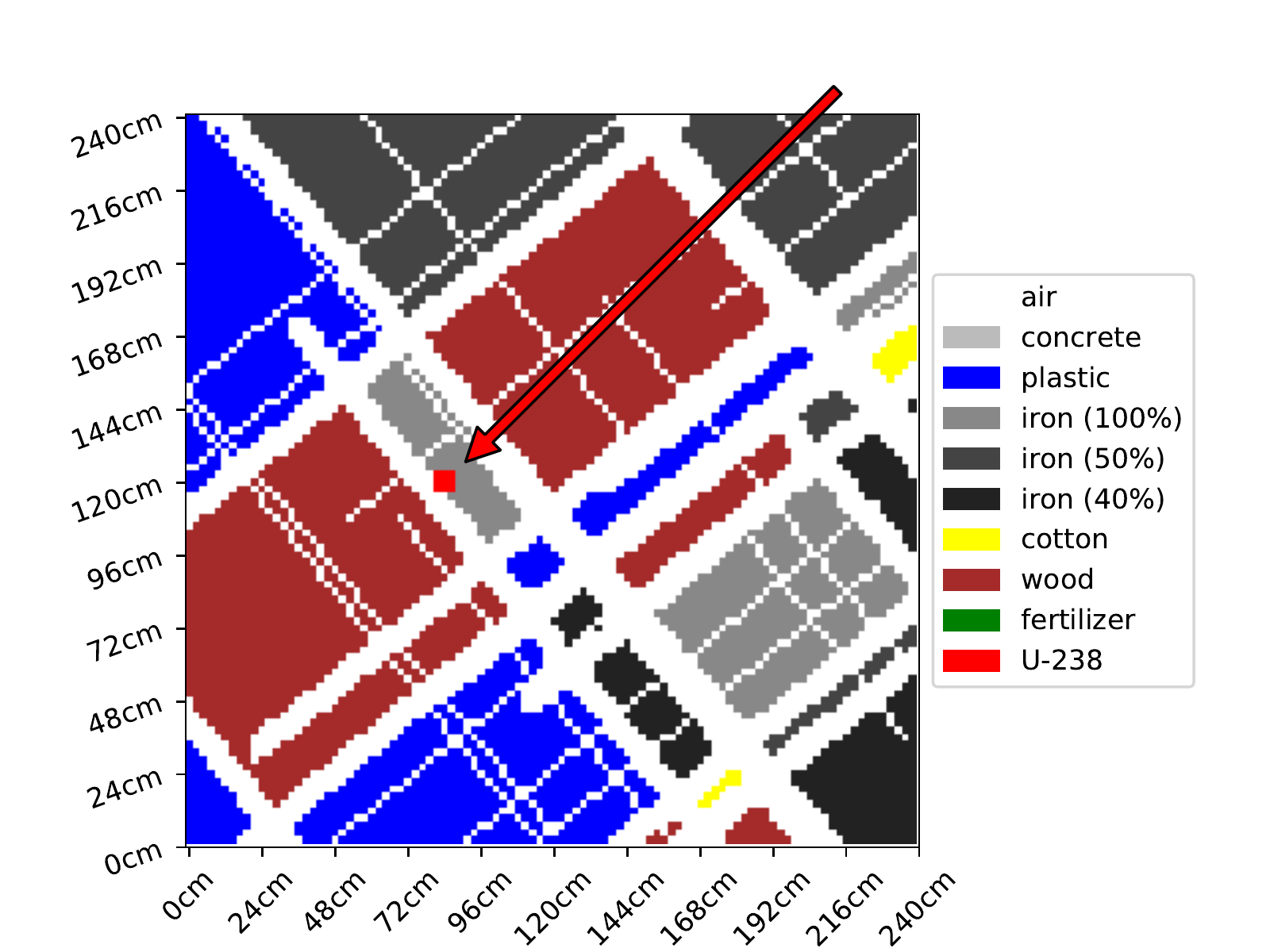}
\end{tabular}
\endgroup
\caption{Left: Backprojection with source detected. Right: Cargo configuration with source configuration indicated by arrow. 100,672 particles detected, 99,683 background particles and 989 source particles. Exposure time is 21.92 seconds.} \label{fig:Ex9}
\end{figure}

\subsection{Performance on Large Scale Dataset} \label{S: Statistical_Performance}
To test the statistical performance of the CNN on a large scale, $1738$ unique cargo configurations are generated using an alternate (to avoid possible inverse crime) generative scheme. For each cargo configuration all four linear detector arrays are present, from zero up to four sources are randomly placed and simulated independently, so that by using linearity of (\ref{eq: Boltzmann}) we can produce $1738\times 16 = 27808$ testing samples.
Particle detections are simulated for an exposure time measured by the expected background detection levels of $20000$, $50000$ and $100000$ particles. The data were fed into the trained CNN for source presence detection.
The results for presence detection are summarized in Table \ref{tab:perf} below.
The  results obtained clearly confirm our expectations (see Section \ref{S:Introduction}).

\begin{table} [h]
\centering
\begin{tabular}{|c|c|c|}
\hline
  Expected Particle Count & Sensitivity & Specificity \\ \hline
100000 & 99.90\%  & 99.71\%  \\ \hline
50000 & 99.78\% & 94.59\%  \\ \hline
20000 & 99.81\% & 36.36\%  \\ \hline
\end{tabular}
\caption{Sensitivity and specificity of the CNN source detection with each source having $1\%$ SNR.} \label{tab:perf}
\end{table}

For comparison, the backprojection showed in the case of around $10^5$ particles sensitivity of about $71\%$ with high specificity (as we have mentioned, the high specificity was built into the backprojection method). In the case of $2\times 10^4$ particles, the sensitivity drops to $52\%$.


We remind the reader that sensitivity, or true positive rate, shows the success of determining the presence of a source (i.e., few false negatives), while specificity reflects how well the absence of the source is detected (i.e., few false positives). High specificity was hardwired into the BP techniques \cite{ADHKK,cargo}, it was only the sensitivity that was questionable.

The accuracy of the prediction generally increases with particle count (and thus observation time), and sufficient particle counts are required for successful detection.
At the low levels (e.g., of 20000 particles and lower) the network seems biased to think that a source is always present. This clearly leads to near $100\%$ sensitivity and an extremely low specificity, which makes the detection practically not feasible, due to high level of false positives.. An explanation could be that the features that are being detected (albeit we do not know what they are) are non-smooth, vs. large smooth background. When the total count is low, the whole dataset becomes non-smooth, which tricks the network.

For $10^5$ particles CNN succeeds extremely well and, as is mentioned above, beats hands down the backprojection technique, which often does not show any statistically significant deviations and thus does not detect presence of the source. Notice that six times higher number of detected particles was required in \cite{ADHKK,X} for stable detection, even without complex cargo being involved. This makes any application of backprojection technique to complex cargo situation groundless and unreliable (even though in almost $71\%$ cases, some of which we already have presented, the location of the source is detected).

\subsection{Number of sources and number of detector arrays}\label{S:Number}
Here we address the question of whether one has to completely surround the object with four detectors, or some results can be achieved with three, two, or one flat detector arrays. We thus have simulated each cargo configuration with zero to four independent sources randomly placed. Just as with the training data, we take several combinations of which sources and detector arrays are present. The combinations are summarized in Table \ref{tab:test_combos} below.
\begin{table}[ht!]
    \centering
    \begin{tabular}{|c|c|c|c|c|c|c|}
    \hline
     \multicolumn{1}{|p{18mm}|}{\centering Number of Sources} & \multicolumn{1}{|p{18mm}|}{\centering One Detector} & \multicolumn{1}{|p{18mm}|}{\centering Two Adjacent Detectors} & \multicolumn{1}{|p{18mm}|}{\centering Two Opposite Detectors} & \multicolumn{1}{|p{18mm}|}{\centering Three Detectors} & \multicolumn{1}{|p{18mm}|}{\centering Four Detectors} & \multicolumn{1}{|p{18mm}|}{\centering Total}\\ \hline
    0 & 6952 & 6952 & 3476 & 6952 & 1738 & 26070 \\ \hline
     1 & 27808 & 27808 & 13904 & 27808 & 6952 & 104280 \\ \hline
     2 & 41712 & 41712 & 20856 & 41712 & 10428 & 156420 \\ \hline
     3 & 27808 & 27808 & 13904 & 27808 & 6952 & 104280 \\ \hline
     4 & 6952 & 6952 & 3476 & 6952 & 1738 & 26070 \\ \hline
     Total & 111232 & 111232 & 55616 & 111232 & 27808 & 417120 \\ \hline
    \end{tabular}
    \caption{Number of testing samples in each category}
    \label{tab:test_combos}
\end{table}

Particle detections are simulated for an exposure time measured by the expected background detection levels of $20000$, $50000$ and $100000$ particles. The data were fed into the trained CNN for inference.
The results for presence detection are summarized in Tables \ref{tab:presence_100k}, \ref{tab:presence_50k} and \ref{tab:presence_20k} and the results for source number prediction are summarized in Tables \ref{tab:number_100k}, \ref{tab:number_50k} and \ref{tab:number_20k}, all in Section \ref{S:Appendix}.

Additionally, we investigated the effect of scaling the strength of each source so that altogether they had the same strength as a single source, thus effectively diluting the localized signature of the source. In the case of backprojection the localized nature of the source is the key justification for the method of \cite{ADHKK}. This would lead one to believe that splitting the source strength will make it more difficult for the CNN to detect any source presence, which is indeed confirmed by the results summarized in Table \ref{tab:perf_ms_2} below.
\begin{table} [h]
\centering
\begin{tabular}{|c|c|c|c|c|c|}
\hline
   Expected Particle Count & \multicolumn{1}{|p{18mm}|}{\centering Sensitivity one source} & \multicolumn{1}{|p{18mm}|}{\centering Sensitivity two sources} & \multicolumn{1}{|p{18mm}|}{\centering Sensitivity three sources} & \multicolumn{1}{|p{18mm}|}{\centering Sensitivity four sources} & \multicolumn{1}{|p{18mm}|}{\centering Specificity}\\ \hline
100000 &  99.74\% & 96.57\%  & 89.67\%  & 82.62\%  & 99.71\%   \\ \hline
50000 & 99.68\%  & 97.25\%  & 93.51\%  & 89.13\%  & 94.59\%  \\ \hline
20000 & 99.99\%  & 99.95\%  & 99.86\%  & 99.77\%  & 36.36\%  \\ \hline

\end{tabular}
\caption{Sensitivity and specificity of the source detection techniques with split source strength.}
\label{tab:perf_ms_2}
\end{table}

\subsection{Observation time} \label{S: Timing}
The above results are presented in terms of the total number of particles detected. The conclusion is natural: the larger - the better. The number of detected particles obviously increases with (essentially proportional to) the time of observation. However, the slope of this increase clearly depends significantly on the type and configuration of the cargo. Thus, the exposure time required to reach a certain level of particle detections is a function of the configuration of the cargo, including source location, material composition, material placement, and background strength. This makes it difficult to predict boundary flux rates, even if the configuration is known, without solving the Boltzmann equation (\ref{eq: Boltzmann}).

\begin{figure}[ht!]
    \centering
    \includegraphics[width=0.8\textwidth]{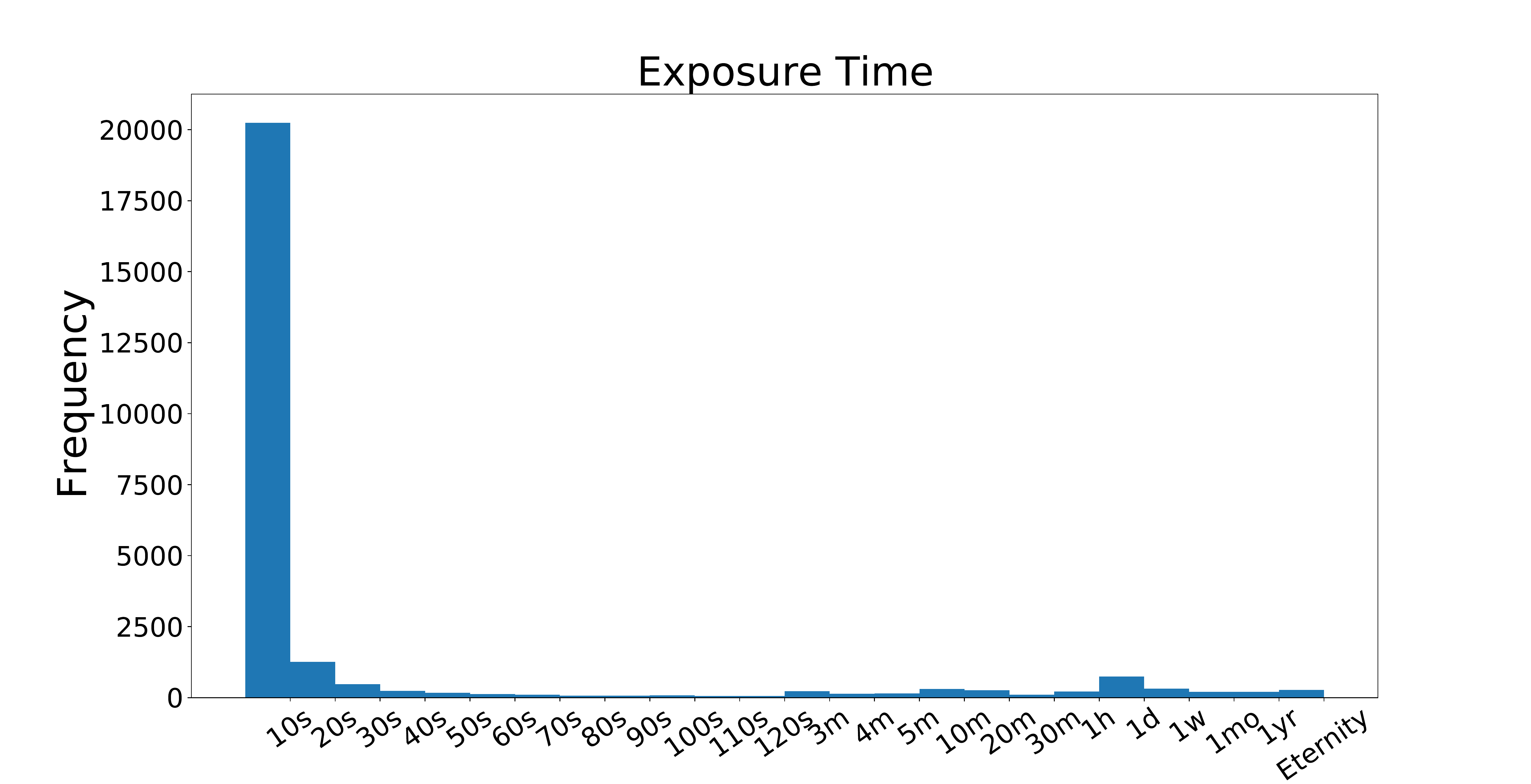}
    \caption{Histogram of the number of runs vs. exposure times required for detection for the testing data set. These times are computed in the case that all four linear detector arrays are present and anywhere between one and four sources are present.}
    \label{fig:ExposureTimes}
\end{figure}

To make a fair numerical experiment, many heavily iron (and thus very shielding) cargo scenarios have been included. Namely, the set of all samples have been divided into 24 sets of equal size, and the probability of choosing iron as the filling of boxes was increasing linearly from zero in the first group to almost one in the 24th one. Figure \ref{fig:ExposureTimes} contains the histogram of the number of runs vs. time required for detection for thousands of configuration runs for detecting the presence of a source emitting on the order of 1000 particles (assuming four detectors). The vast majority would require time measured in seconds.

Generally speaking, one would epect the large bin on the left-hand side to correspond to configurations with less high-Z materials, and the larger bins on the right-hand side correspond to configurations with more high-Z materials. It can certainly become unrealistic to detect many source particles in some of the latter cases. Nevertheless, as is evidenced by some of the examples presented, as well as statistics presented in Section \ref{S: Statistical_Performance}, quite a few  configurations of high-Z material exist where presence of source(s) source can be detected in a reasonable amount of time. These lower exposure time scenarios would be the most appropriate cases for detecting illicit nuclear materials at border crossings. Some of the longer exposure times (on the order of several minutes to perhaps several days) would be appropriate for detection of illicit nuclear materials in shipping containers on cargo ships, where scanning can be done while the container is in transit.

Additionally, it is important to note that if one restricts oneself to a smaller number of detector arrays (incomplete view), it will take longer to reach the same exposure level and thus would add to the number of undetected cases.

\section{Remarks and Conclusions}\label{S:remarks}

\begin{itemize}
\item Our work shows that the deep learning approach significantly improves over detection by backprojection techniques of \cite{TKK,cargo,X} and works for complex attenuating and scattering cargo scenarios, where the latter fails completely. This confirms the opinion expressed in \cite{cargo} that some information about source presence was there.
\item This article concentrates on the cases of presence of complex cargo and much (an order of magnitude) lower number of $\gamma$-photon detected than in \cite{ADHKK,X}. This makes the backprojection detection algorithm of these works not only weak, but also groundless.
\item The network performs well detecting the number of up to four sources (although naturally somewhat less successfully than detecting mere presence of a source).
\item The authors want to make clear that when producing the results of this paper, no processing (e.g., backprojecting) of the raw detector data is done before feeding it to the network. Since the authors do not know what features would be of importance, we have decided to not impose our prejudices on the data (especially taking into account that backprojection is a smoothing operator, and the relevant information is most probably contained in some sharper features).
\item The exposure time required for detection is discussed in Section \ref{S: Timing}. The histogram in Figure \ref{fig:ExposureTimes} shows detection in a matter of second for a vast majority of configurations.

It is clear that there are some unbeatable shieldings, so one cannot aim for the $100\%$ success rate. In such cases, other detection techniques could be used: from methods of detecting presence of significant amounts of (shielding) high-Z materials, to neutron emission detection, to human intelligence.
\item A strong effort has been made to avoid committing an inverse crime. The testing samples have been produced by an algorithm independent of the one used for the training data. The testing cargo geometries were different from the ones not encountered in the training data, so there was no intersection between the two data-sets.
\item A variety of symmetry rules, including rotational symmetry and mirror symmetry were applied randomly to some of the configurations and their material content, to check whether presence or absence of the symmetry influence the detectability. The network performance does not seem to react to this.
\item The reader should not think that retraining was needed for different tasks and situations, e.g. for heavy iron cargo, or for detecting the number of sources, rather their mere presence. This all was done with a single trained network.
\item Four planar Compton detectors forming a square surrounding the object of interrogation were assumed. It seems that this is the most practical design of such detectors. Effects of removal of some of the detectors have also been studied (Section \ref{S:Number}).

    The rectangular shape causes some problems, though, e.g. in backprojection method they create (easily removable) corner artifacts. More importantly, this design lacks full rotational invariance, which could be beneficial for the NN design.
    On the other hand, the rectangular case is challenged by appearance of tilted cargo structures in the test samples, while they were absent in the training data. The network, however, clearly has overcome this difficulty.
\item There are various further improvements that one should attempt (and are being attempted). Some of them are addressed below.
\begin{enumerate}
\item It would have been great to figure out what specifically were the signs of presence of the source that the network has learned. This would open a door for developing more analytic methods. However, at this moment the authors do not know what these features are.
\item Producing many more training data is a serious stumbling block in 2D, and especially in 3D case.
\item The CNN architecture should be improved, aiming to reach shorter observation time and even lower SNR levels.
\item We are working on moving to the more realistic 3D situation.  The significant difference here is, first, the much higher dimensionality of the data (5D) and corresponding much more massive computations that are needed. Second, in 3D, unlike 2D (where a cone consists just of two rays), the Compton data differ significantly from the usual Radon ones. In particular, an issue arises of how to bin the five-dimensional Compton data in such a way, that the use of CNN could be warranted.
\item The neural network (NN) approach should be tested on real data, which the authors clearly do not have. However, the radiative transport forward computations we used are commonly practiced in nuclear engineering, seem to be very realistic, and involve realistic material parameters. There is a chance that when novel neutron detectors that are being developed are deployed, we could get some real data.
\item The approach we describe indicates presence of a source, but not its location (at least in the heavy iron cargo case). One wonders whether location can also be attempted.
\item Although the results presented have been obtained by the same once trained NN, during research various designs of the NN and training sets have been experimented with, all showing consistent ability of detection. It would be still important to study further the model uncertainty (e.g., by using the dropout technique \cite{Unc}). This will be done in a future work.

    Meanwhile, although the testing samples often deviated from the structures used in the training set, our results have shown that the NN generalized extremely well. High experimental levels of the sensitivity and specificity, as well as more detailed information presented in Section \ref{S: Statistical_Performance} about statistical spread of the results instill confidence in the suitability of the network as a detection tool.

    The imperfection of Compton camera detections has been partially addressed by randomizing the source strength and location and finite bin sizes for the detected data. Depending on the quality of the future detectors, the bin sizes might have to be increased and new study conducted.
\item When source particles scatter they lose energy. If source particles downscatter to lower energy groups we will lose them in our data since we only use the highest energy group. It would be interesting to try and use these lower energy groups in either a 3D convolution with 1 channel or a 2D convolution with multiple channels to see if we can get better results.
\end{enumerate}
\end{itemize}

\section{Acknowledgements} \label{S:acknowledgements}
The first two authors acknowledge the support from the National Science Foundation through the DMS grant \#1816430 and Texas A\&M Cyclotron Institute. The third author has been partially supported through a grant by the Department of the Defense, Defense Threat Reduction Agency under Award No. HDTRA1-18-1-0020. The content of the article does not necessarily reflect the position or the policy of the federal government, and no official endorsement should be inferred.

The authors are truly indebted to the three referees, whose detailed comments served not only to improve exposition, but even more importantly, attracted our attention to some features we have missed.

\section*{References}


\newpage
\section{Appendix} \label{S:Appendix}

\subsection{Algorithm for Procedural Generation of Cargo Configurations}\label{SS:algorithm}
\begin{algorithm}[ht!]
\SetAlgoLined
 Generate Perlin noise in cargo\;
 Initialize $n_x$ and $n_y$ to desired number of vertical and horizontal boundaries (numbers can be chosen randomly)\;
 Sum Perlin noise over rows and columns to produce noise function on edge of cargo\;
 Randomly select $n_x$ distinct $x-$coordinates for vertical boundaries and $n_y$ distinct $y-$coordinates for horizontal boundaries according to edge noise functions. Store in $x$ and $y$ respectively.\;
 $n_{iter}=0$ \;
 \While{$n_x > 0$ or $n_y > 0$}{
  \uIf{$n_{iter}$ is even and $n_x > 0$}{
    Determine all existing boundary points along the line $(x[n_{iter}/2],y)$.
    Randomly select a starting point $y_s$ and ending point $y_e$ from among the existing boundary points according to previously generated Perlin noise. Set all points between $(x[n_{iter}/2],y_s)$ and $(x[n_{iter}/2],y_e)$ to boundary points.\;
   $n_x=n_x-1$\;
   $n_{iter}=n_{iter}+1$\;
   }
   \uElseIf{$n_{iter}$ is odd and $n_y > 0$}{
   Determine all existing boundary points along the line $(x,(y[(n_{iter}-1)/2])$.
    Randomly select a starting point $x_s$ and ending point $x_e$ from among the existing boundary points according to previously generated Perlin noise. Set all points between $(x_s,(y[(n_{iter}-1)/2])$ and $(x_e,(y[(n_{iter}-1)/2])$ to boundary points.\;
   $n_y=n_y-1$\;
   $n_{iter}=n_{iter}+1$\;
  }
 }
 Identify connected components (Scipy.Measure.Label)\;
 \uIf{Rotational Symmetry Desired}{
 Copy one quadrant of the configuration over all others with appropriate rotation\;
 }
 \uIf{Mirror Symmetry Desired}{
 Copy one side of the configuration over the other with mirroring \;
 }
 ...
 Randomly assign material identification to each connected component \;
 Save configuration to file\;
 \KwResult{Single cargo configuration }
 \caption{Procedural Cargo Configuration}
\end{algorithm}
\subsection{Statistics of the source(s) detection}

\begin{table}[h]
  \begin{subtable}[t]{.33\linewidth}%
    \centering%
    \resizebox{\columnwidth}{!}{
\begin{tabular}{|p{20mm}|p{20mm}|p{20mm}|}
\hline
  \centering Number of Sources &  \multicolumn{2}{|c|}{\vspace{-5mm} CNN Prediction} \\ \cline{2-3}
                    &  \centering Source Present & \centering No Source Present \cr \hline
  \centering 0 & \centering 6.04 \% & \centering 93.36 \% \cr \hline
  \centering 1 & \centering 98.35 \% & \centering 1.65 \% \cr \hline
  \centering 2 & \centering 99.90 \% & \centering 0.10 \% \cr \hline
  \centering 3 & \centering 99.98 \% & \centering 0.02 \% \cr \hline
  \centering 4 & \centering 100.00 \% & \centering 0.00 \% \cr \hline
\end{tabular}} \caption{}
  \end{subtable}%
  \begin{subtable}[t]{.33\linewidth}
   \centering%
   \resizebox{\columnwidth}{!}{
\begin{tabular}{|p{20mm}|p{20mm}|p{20mm}|}
\hline
  \centering Number of Sources &  \multicolumn{2}{|c|}{\vspace{-5mm} CNN Prediction} \\ \cline{2-3}
                    &  \centering Source Present & \centering No Source Present \cr \hline
  \centering 0 & \centering 3.31 \% & \centering 96.69 \% \cr \hline
  \centering 1 & \centering 99.20 \% & \centering 0.80 \% \cr \hline
  \centering 2 & \centering 99.96 \% & \centering 0.04 \% \cr \hline
  \centering 3 & \centering 100.00 \% & \centering 0.00 \% \cr \hline
  \centering 4 & \centering 100.00 \% & \centering 0.00 \% \cr \hline
\end{tabular}} \caption{}
  \end{subtable}
  \begin{subtable}[t]{.33\linewidth}%
    \centering%
    \resizebox{\columnwidth}{!}{
\begin{tabular}{|p{20mm}|p{20mm}|p{20mm}|}
\hline
  \centering Number of Sources &  \multicolumn{2}{|c|}{\vspace{-5mm} CNN Prediction} \\ \cline{2-3}
                    &  \centering Source Present & \centering No Source Present \cr \hline
  \centering 0 & \centering 1.41 \% & \centering 98.59 \% \cr \hline
  \centering 1 & \centering 99.15 \% & \centering 0.85 \% \cr \hline
  \centering 2 & \centering 99.96 \% & \centering 0.04 \% \cr \hline
  \centering 3 & \centering 100.00 \% & \centering 0.00 \% \cr \hline
  \centering 4 & \centering 100.00 \% & \centering 0.00 \% \cr \hline
\end{tabular}} \caption{}
  \end{subtable}\par \bigskip
  \begin{subtable}[t]{.33\linewidth}
   \centering%
   \resizebox{\columnwidth}{!}{
\begin{tabular}{|p{20mm}|p{20mm}|p{20mm}|}
\hline
  \centering Number of Sources &  \multicolumn{2}{|c|}{\vspace{-5mm} CNN Prediction} \\ \cline{2-3}
                    &  \centering Source Present & \centering No Source Present \cr \hline
  \centering 0 & \centering 0.55 \% & \centering 99.45 \% \cr \hline
  \centering 1 & \centering 99.19 \% & \centering 0.81 \% \cr \hline
  \centering 2 & \centering 99.97 \% & \centering 0.03 \% \cr \hline
  \centering 3 & \centering 99.99 \% & \centering 0.01 \% \cr \hline
  \centering 4 & \centering 99.99 \% & \centering 0.01 \% \cr \hline
\end{tabular}} \caption{}
  \end{subtable}
  \begin{subtable}[t]{.33\linewidth}%
    \centering%
    \resizebox{\columnwidth}{!}{
\begin{tabular}{|p{20mm}|p{20mm}|p{20mm}|}
\hline
  \centering Number of Sources &  \multicolumn{2}{|c|}{\vspace{-5mm} CNN Prediction} \\ \cline{2-3}
                    &  \centering Source Present & \centering No Source Present \cr \hline
  \centering 0 & \centering 0.29 \% & \centering 99.71 \% \cr \hline
  \centering 1 & \centering 99.27 \% & \centering 0.73 \% \cr \hline
  \centering 2 & \centering 100.00 \% & \centering 0.00 \% \cr \hline
  \centering 3 & \centering 100.00 \% & \centering 0.00 \% \cr \hline
  \centering 4 & \centering 100.00 \% & \centering 0.00 \% \cr \hline
\end{tabular}} \caption{}
  \end{subtable}%
  \begin{subtable}[t]{.33\linewidth}
   \centering%
   \resizebox{\columnwidth}{!}{
\begin{tabular}{|p{20mm}|p{20mm}|p{20mm}|}
\hline
  \centering Number of Sources &  \multicolumn{2}{|c|}{\vspace{-5mm} CNN Prediction} \\ \cline{2-3}
                    &  \centering Source Present & \centering No Source Present \cr \hline
  \centering 0 & \centering 2.85 \% & \centering 97.15 \% \cr \hline
  \centering 1 & \centering 98.97 \% & \centering 1.03 \% \cr \hline
  \centering 2 & \centering 99.95 \% & \centering 0.05 \% \cr \hline
  \centering 3 & \centering 99.99 \% & \centering 0.01 \% \cr \hline
  \centering 4 & \centering 100.00 \% & \centering 0.00 \% \cr \hline
\end{tabular}} \caption{}
  \end{subtable}
  \caption{Source Presence Prediction Performance at 100,000 background particle level. (a): One Linear Detector Array, (b): Two Adjacent Linear Detector Arrays, (c): Two Opposite Linear Detector Arrays, (d): Three Linear Detector Arrays, (e): Four Linear Detector Arrays, (f): Performance Across Whole Data Set (Every combination of detectors)}\label{tab:presence_100k}
\end{table}

\begin{table}
  \begin{subtable}[t]{.33\linewidth}%
    \centering%
    \resizebox{\columnwidth}{!}{
\begin{tabular}{|p{20mm}|p{20mm}|p{20mm}|}
\hline
  \centering Number of Sources &  \multicolumn{2}{|c|}{\vspace{-5mm} CNN Prediction} \\ \cline{2-3}
                    &  \centering Source Present & \centering No Source Present \cr \hline
  \centering 0 & \centering 15.03 \% & \centering 84.97 \% \cr \hline
  \centering 1 & \centering 97.51 \% & \centering 2.49 \% \cr \hline
  \centering 2 & \centering 99.76 \% & \centering 0.24 \% \cr \hline
  \centering 3 & \centering 99.98 \% & \centering 0.02 \% \cr \hline
  \centering 4 & \centering 100.00 \% & \centering 0.00 \% \cr \hline
\end{tabular}} \caption{}
  \end{subtable}%
  \begin{subtable}[t]{.33\linewidth}
   \centering%
   \resizebox{\columnwidth}{!}{
\begin{tabular}{|p{20mm}|p{20mm}|p{20mm}|}
\hline
  \centering Number of Sources &  \multicolumn{2}{|c|}{\vspace{-5mm} CNN Prediction} \\ \cline{2-3}
                    &  \centering Source Present & \centering No Source Present \cr \hline
  \centering 0 & \centering 11.39 \% & \centering 88.81 \% \cr \hline
  \centering 1 & \centering 98.86 \% & \centering 1.14 \% \cr \hline
  \centering 2 & \centering 99.93 \% & \centering 0.07 \% \cr \hline
  \centering 3 & \centering 100.00 \% & \centering 0.00 \% \cr \hline
  \centering 4 & \centering 100.00 \% & \centering 0.00 \% \cr \hline
\end{tabular}} \caption{}
  \end{subtable}
  \begin{subtable}[t]{.33\linewidth}%
    \centering%
    \resizebox{\columnwidth}{!}{
\begin{tabular}{|p{20mm}|p{20mm}|p{20mm}|}
\hline
  \centering Number of Sources &  \multicolumn{2}{|c|}{\vspace{-5mm} CNN Prediction} \\ \cline{2-3}
                    &  \centering Source Present & \centering No Source Present \cr \hline
  \centering 0 & \centering 7.91 \% & \centering 92.09 \% \cr \hline
  \centering 1 & \centering 99.04 \% & \centering 0.96 \% \cr \hline
  \centering 2 & \centering 99.95 \% & \centering 0.05 \% \cr \hline
  \centering 3 & \centering 99.99 \% & \centering 0.01 \% \cr \hline
  \centering 4 & \centering 100.00 \% & \centering 0.00 \% \cr \hline
\end{tabular}} \caption{}
  \end{subtable}\par \bigskip
  \begin{subtable}[t]{.33\linewidth}
   \centering%
   \resizebox{\columnwidth}{!}{
\begin{tabular}{|p{20mm}|p{20mm}|p{20mm}|}
\hline
  \centering Number of Sources &  \multicolumn{2}{|c|}{\vspace{-5mm} CNN Prediction} \\ \cline{2-3}
                    &  \centering Source Present & \centering No Source Present \cr \hline
  \centering 0 & \centering 6.67 \% & \centering 93.33 \% \cr \hline
  \centering 1 & \centering 99.12 \% & \centering 0.87 \% \cr \hline
  \centering 2 & \centering 99.96 \% & \centering 0.04 \% \cr \hline
  \centering 3 & \centering 99.99 \% & \centering 0.01 \% \cr \hline
  \centering 4 & \centering 99.99 \% & \centering 0.01 \% \cr \hline
\end{tabular}} \caption{}
  \end{subtable}
  \begin{subtable}[t]{.33\linewidth}%
    \centering%
    \resizebox{\columnwidth}{!}{
\begin{tabular}{|p{20mm}|p{20mm}|p{20mm}|}
\hline
  \centering Number of Sources &  \multicolumn{2}{|c|}{\vspace{-5mm} CNN Prediction} \\ \cline{2-3}
                    &  \centering Source Present & \centering No Source Present \cr \hline
  \centering 0 & \centering 5.41 \% & \centering 94.59 \% \cr \hline
  \centering 1 & \centering 99.18 \% & \centering 0.82 \% \cr \hline
  \centering 2 & \centering 99.99 \% & \centering 0.01 \% \cr \hline
  \centering 3 & \centering 100.00 \% & \centering 0.00 \% \cr \hline
  \centering 4 & \centering 100.00 \% & \centering 0.00 \% \cr \hline
\end{tabular}} \caption{}
  \end{subtable}%
  \begin{subtable}[t]{.33\linewidth}
   \centering%
   \resizebox{\columnwidth}{!}{
\begin{tabular}{|p{20mm}|p{20mm}|p{20mm}|}
\hline
  \centering Number of Sources &  \multicolumn{2}{|c|}{\vspace{-5mm} CNN Prediction} \\ \cline{2-3}
                    &  \centering Source Present & \centering No Source Present \cr \hline
  \centering 0 & \centering 10.24 \% & \centering 89.76 \% \cr \hline
  \centering 1 & \centering 98.62 \% & \centering 1.38 \% \cr \hline
  \centering 2 & \centering 99.90 \% & \centering 0.10 \% \cr \hline
  \centering 3 & \centering 99.99 \% & \centering 0.01 \% \cr \hline
  \centering 4 & \centering 100.00 \% & \centering 0.00 \% \cr \hline
\end{tabular}} \caption{}
  \end{subtable}
  \caption{Source Presence Prediction Performance at 50,000 background particle level. (a): One Linear Detector Array, (b): Two Adjacent Linear Detector Arrays, (c): Two Opposite Linear Detector Arrays, (d): Three Linear Detector Arrays, (e): Four Linear Detector Arrays, (f): Performance Across Whole Data Set (Every combination of detectors)}\label{tab:presence_50k}
\end{table}

\begin{table}
  \begin{subtable}[t]{.33\linewidth}%
    \centering%
    \resizebox{\columnwidth}{!}{
\begin{tabular}{|p{20mm}|p{20mm}|p{20mm}|}
\hline
  \centering Number of Sources &  \multicolumn{2}{|c|}{\vspace{-5mm} CNN Prediction} \\ \cline{2-3}
                    &  \centering Source Present & \centering No Source Present \cr \hline
  \centering 0 & \centering 66.59 \% & \centering 33.41 \% \cr \hline
  \centering 1 & \centering 97.77 \% & \centering 2.23 \% \cr \hline
  \centering 2 & \centering 99.57 \% & \centering 0.43 \% \cr \hline
  \centering 3 & \centering 99.90 \% & \centering 0.10 \% \cr \hline
  \centering 4 & \centering 100.00 \% & \centering 0.00 \% \cr \hline
\end{tabular}} \caption{}
  \end{subtable}%
  \begin{subtable}[t]{.33\linewidth}
   \centering%
   \resizebox{\columnwidth}{!}{
\begin{tabular}{|p{20mm}|p{20mm}|p{20mm}|}
\hline
  \centering Number of Sources &  \multicolumn{2}{|c|}{\vspace{-5mm} CNN Prediction} \\ \cline{2-3}
                    &  \centering Source Present & \centering No Source Present \cr \hline
  \centering 0 & \centering 66.20 \% & \centering 33.80 \% \cr \hline
  \centering 1 & \centering 98.90 \% & \centering 1.10 \% \cr \hline
  \centering 2 & \centering 99.86 \% & \centering 0.14 \% \cr \hline
  \centering 3 & \centering 99.99 \% & \centering 0.01 \% \cr \hline
  \centering 4 & \centering 100.00 \% & \centering 0.00 \% \cr \hline
\end{tabular}} \caption{}
  \end{subtable}
  \begin{subtable}[t]{.33\linewidth}%
    \centering%
    \resizebox{\columnwidth}{!}{
\begin{tabular}{|p{20mm}|p{20mm}|p{20mm}|}
\hline
 \centering Number of Sources &  \multicolumn{2}{|c|}{\vspace{-5mm} CNN Prediction} \\ \cline{2-3}
                    &  \centering Source Present & \centering No Source Present \cr \hline
  \centering 0 & \centering 64.79 \% & \centering 35.21 \% \cr \hline
  \centering 1 & \centering 99.17 \% & \centering 0.83 \% \cr \hline
  \centering 2 & \centering 99.91 \% & \centering 0.09 \% \cr \hline
  \centering 3 & \centering 99.95 \% & \centering 0.05 \% \cr \hline
  \centering 4 & \centering 100.00 \% & \centering 0.00 \% \cr \hline
\end{tabular}} \caption{}
  \end{subtable}\par \bigskip
  \begin{subtable}[t]{.33\linewidth}
   \centering%
   \resizebox{\columnwidth}{!}{
\begin{tabular}{|p{20mm}|p{20mm}|p{20mm}|}
\hline
  \centering Number of Sources &  \multicolumn{2}{|c|}{\vspace{-5mm} CNN Prediction} \\ \cline{2-3}
                    &  \centering Source Present & \centering No Source Present \cr \hline
  \centering 0 & \centering 62.87 \% & \centering 37.13 \% \cr \hline
  \centering 1 & \centering 99.32 \% & \centering 0.68 \% \cr \hline
  \centering 2 & \centering 99.94 \% & \centering 0.06 \% \cr \hline
  \centering 3 & \centering 99.99 \% & \centering 0.01 \% \cr \hline
  \centering 4 & \centering 99.99 \% & \centering 0.01 \% \cr \hline
\end{tabular}} \caption{}
  \end{subtable}
  \begin{subtable}[t]{.33\linewidth}%
    \centering%
    \resizebox{\columnwidth}{!}{
\begin{tabular}{|p{20mm}|p{20mm}|p{20mm}|}
\hline
  \centering Number of Sources &  \multicolumn{2}{|c|}{\vspace{-5mm} CNN Prediction} \\ \cline{2-3}
                    &  \centering Source Present & \centering No Source Present \cr \hline
  \centering 0 & \centering 63.64 \% & \centering 36.36 \% \cr \hline
  \centering 1 & \centering 99.34 \% & \centering 0.66 \% \cr \hline
  \centering 2 & \centering 99.99 \% & \centering 0.01 \% \cr \hline
  \centering 3 & \centering 100.00 \% & \centering 0.00 \% \cr \hline
  \centering 4 & \centering 100.00 \% & \centering 0.00 \% \cr \hline
\end{tabular}} \caption{}
  \end{subtable}%
  \begin{subtable}[t]{.33\linewidth}
   \centering%
   \resizebox{\columnwidth}{!}{
\begin{tabular}{|p{20mm}|p{20mm}|p{20mm}|}
\hline
  \centering Number of Sources &  \multicolumn{2}{|c|}{\vspace{-5mm} CNN Prediction} \\ \cline{2-3}
                    &  \centering Source Present & \centering No Source Present \cr \hline
  \centering 0 & \centering 65.06 \% & \centering 34.94 \% \cr \hline
  \centering 1 & \centering 98.77 \% & \centering 1.23 \% \cr \hline
  \centering 2 & \centering 99.82 \% & \centering 0.18 \% \cr \hline
  \centering 3 & \centering 99.96 \% & \centering 0.04 \% \cr \hline
  \centering 4 & \centering 100.00 \% & \centering 0.00 \% \cr \hline
\end{tabular}} \caption{}
  \end{subtable}
  \caption{Source Presence Prediction Performance at 20,000 background particle level. (a): One Linear Detector Array, (b): Two Adjacent Linear Detector Arrays, (c): Two Opposite Linear Detector Arrays, (d): Three Linear Detector Arrays, (e): Four Linear Detector Arrays, (f): Performance Across Whole Data Set (Every combination of detectors)}\label{tab:presence_20k}
\end{table}

\begin{table}

  \begin{subtable}[t]{.5\linewidth}%
    \centering%
\resizebox{\columnwidth}{!}{
\begin{tabular}{|p{12mm}|c|c|c|c|c|}
\hline
   \centering No. Sources &  \multicolumn{5}{|c|}{\vspace{-5mm}CNN Prediction}\\ \cline{2-6}
   &  \centering 0 & \centering 1 & \centering 2 & \centering 3 & \centering 4 \cr \hline
  \centering 0 & \centering 94.88 \% & \centering 3.90 \% & \centering 0.81 \% & 0.27 \% & 0.14 \% \cr \hline
  \centering 1 & \centering 1.73 \% & \centering 76.00 \% & \centering 20.05 \% & 1.87 \% & 0.35 \% \cr \hline
  \centering 2 & \centering 0.11 \% & \centering 6.93 \% & \centering 71.89 \% & 20.04 \% & 1.03 \% \cr \hline
  \centering 3 & \centering 0.01 \% & \centering 0.54 \% & \centering 22.46 \% & 71.03 \% & 5.96 \% \cr \hline
  \centering 4 & \centering 0.00 \% & \centering 0.06 \% & \centering 4.00 \% & 62.62 \% & 33.32 \% \cr \hline
\end{tabular}} \caption{}
  \end{subtable}
  \begin{subtable}[t]{.5\linewidth}%
    \centering%
    \resizebox{\columnwidth}{!}{
\begin{tabular}{|p{12mm}|c|c|c|c|c|}
\hline
   \centering No. Sources &  \multicolumn{5}{|c|}{\vspace{-5mm}CNN Prediction}\\ \cline{2-6}
   &  \centering 0 & \centering 1 & \centering 2 & \centering 3 & \centering 4 \cr \hline
  \centering 0 & \centering 97.41 \% & \centering 2.29 \% & \centering 0.26 \% & 0.04 \% & 0.00 \% \cr \hline
  \centering 1 & \centering 0.88 \% & \centering 78.80 \% & \centering 19.75 \% & 0.55 \% & 0.02 \% \cr \hline
  \centering 2 & \centering 0.05 \% & \centering 4.41 \% & \centering 76.05 \% & 19.38 \% & 0.11 \% \cr \hline
  \centering 3 & \centering 0.00 \% & \centering 0.33 \% & \centering 17.19 \% & 77.73 \% & 4.75 \% \cr \hline
  \centering 4 & \centering 0.00 \% & \centering 0.10 \% & \centering 2.11 \% & 59.74 \% & 38.05 \% \cr \hline
\end{tabular}} \caption{}
  \end{subtable}\par\bigskip
  \begin{subtable}[t]{.5\linewidth}%
    \centering%
\resizebox{\columnwidth}{!}{
\begin{tabular}{|p{12mm}|c|c|c|c|c|}
\hline
   \centering No. Sources &  \multicolumn{5}{|c|}{\vspace{-5mm}CNN Prediction}\\ \cline{2-6}
   &  \centering 0 & \centering 1 & \centering 2 & \centering 3 & \centering 4 \cr \hline
  \centering 0 & \centering 98.85 \% & \centering 1.04 \% & \centering 0.12 \% & 0.00 \% & 0.00 \% \cr \hline
  \centering 1 & \centering 0.87 \% & \centering 80.48 \% & \centering 18.46 \% & 0.19 \% & 0.00 \% \cr \hline
  \centering 2 & \centering 0.04 \% & \centering 4.38 \% & \centering 76.09 \% & 19.41 \% & 0.08 \% \cr \hline
  \centering 3 & \centering 0.00 \% & \centering 0.40 \% & \centering 15.94 \% & 78.41 \% & 5.25 \% \cr \hline
  \centering 4 & \centering 0.00 \% & \centering 0.06 \% & \centering 2.68 \% & 58.03 \% & 39.24 \% \cr \hline
\end{tabular}} \caption{}
  \end{subtable}
  \begin{subtable}[t]{.5\linewidth}%
    \centering%
    \resizebox{\columnwidth}{!}{
\begin{tabular}{|p{12mm}|c|c|c|c|c|}
\hline
   \centering No. Sources &  \multicolumn{5}{|c|}{\vspace{-5mm}CNN Prediction}\\ \cline{2-6}
   &  \centering 0 & \centering 1 & \centering 2 & \centering 3 & \centering 4 \cr \hline
  \centering 0 & \centering 99.73 \% & \centering 0.26 \% & \centering 0.01 \% & 0.00 \% & 0.00 \% \cr \hline
  \centering 1 & \centering 0.82 \% & \centering 79.06 \% & \centering 20.06 \% & 0.07 \% & 0.00 \% \cr \hline
  \centering 2 & \centering 0.03 \% & \centering 3.62 \% & \centering 75.65 \% & 20.61 \% & 0.09 \% \cr \hline
  \centering 3 & \centering 0.01 \% & \centering 0.37 \% & \centering 15.09 \% & 78.96 \% & 5.57 \% \cr \hline
  \centering 4 & \centering 0.01 \% & \centering 0.14 \% & \centering 2.20 \% & 57.95 \% & 39.69 \% \cr \hline
\end{tabular}} \caption{}
  \end{subtable}\par\bigskip
  \begin{subtable}[t]{.5\linewidth}%
    \centering%
\resizebox{\columnwidth}{!}{
\begin{tabular}{|p{12mm}|c|c|c|c|c|}
\hline
   \centering No. Sources &  \multicolumn{5}{|c|}{\vspace{-5mm}CNN Prediction}\\ \cline{2-6}
   &  \centering 0 & \centering 1 & \centering 2 & \centering 3 & \centering 4 \cr \hline
  \centering 0 & \centering 99.83 \% & \centering 0.17 \% & \centering 0.00 \% & 0.00 \% & 0.00 \% \cr \hline
  \centering 1 & \centering 0.76 \% & \centering 78.91 \% & \centering 20.24 \% & 0.09 \% & 0.00 \% \cr \hline
  \centering 2 & \centering 0.00 \% & \centering 3.10 \% & \centering 75.84 \% & 20.94 \% & 0.12 \% \cr \hline
  \centering 3 & \centering 0.00 \% & \centering 0.13 \% & \centering 11.84 \% & 82.39 \% & 5.64 \% \cr \hline
  \centering 4 & \centering 0.00 \% & \centering 0.00 \% & \centering 1.27 \% & 58.63 \% & 40.10 \% \cr \hline
\end{tabular}} \caption{}
  \end{subtable}
  \begin{subtable}[t]{.5\linewidth}%
    \centering%
    \resizebox{\columnwidth}{!}{
\begin{tabular}{|p{12mm}|c|c|c|c|c|}
\hline
   \centering No. Sources &  \multicolumn{5}{|c|}{\vspace{-5mm}CNN Prediction}\\ \cline{2-6}
   &  \centering 0 & \centering 1 & \centering 2 & \centering 3 & \centering 4 \cr \hline
  \centering 0 & \centering 97.71 \% & \centering 1.87 \% & \centering 0.30 \% & 0.08 \% & 0.04 \% \cr \hline
  \centering 1 & \centering 1.08 \% & \centering 78.36 \% & \centering 19.77 \% & 0.70 \% & 0.10 \% \cr \hline
  \centering 2 & \centering 0.05 \% & \centering 4.78 \% & \centering 74.83 \% & 19.99 \% & 0.35 \% \cr \hline
  \centering 3 & \centering 0.01 \% & \centering 0.39 \% & \centering 17.51 \% & 76.67 \% & 5.42 \% \cr \hline
  \centering 4 & \centering 0.00 \% & \centering 0.09 \% & \centering 2.66 \% & 59.73 \% & 37.52 \% \cr \hline
\end{tabular}} \caption{}
  \end{subtable}
  \caption{Source Number Prediction Performance at 100,000 background particle level. (a): One Linear Detector Array, (b): Two Adjacent Linear Detector Arrays, (c): Two Opposite Linear Detector Arrays, (d): Three Linear Detector Arrays, (e): Four Linear Detector Arrays, (f): Performance Across Whole Data Set (Every combination of detectors)}\label{tab:number_100k}
\end{table}

\begin{table}

  \begin{subtable}[t]{.5\linewidth}%
    \centering%
\resizebox{\columnwidth}{!}{
\begin{tabular}{|p{12mm}|c|c|c|c|c|}
\hline
   \centering No. Sources &  \multicolumn{5}{|c|}{\vspace{-5mm}CNN Prediction}\\ \cline{2-6}
   &  \centering 0 & \centering 1 & \centering 2 & \centering 3 & \centering 4 \cr \hline
  \centering 0 & \centering 86.12 \% & \centering 12.15 \% & \centering 1.18 \% & 0.40 \% & 0.14 \% \cr \hline
  \centering 1 & \centering 2.40 \% & \centering 78.08 \% & \centering 17.33 \% & 1.82 \% & 0.37 \% \cr \hline
  \centering 2 & \centering 0.22 \% & \centering 14.08 \% & \centering 72.57 \% & 12.27 \% & 0.87 \% \cr \hline
  \centering 3 & \centering 0.03 \% & \centering 1.52 \% & \centering 39.20 \% & 56.29 \% & 2.96 \% \cr \hline
  \centering 4 & \centering 0.00 \% & \centering 0.24 \% & \centering 9.34 \% & 74.58 \% & 15.84 \% \cr \hline
\end{tabular}} \caption{}
  \end{subtable}
  \begin{subtable}[t]{.5\linewidth}%
    \centering%
    \resizebox{\columnwidth}{!}{
\begin{tabular}{|p{12mm}|c|c|c|c|c|}
\hline
   \centering No. Sources &  \multicolumn{5}{|c|}{\vspace{-5mm}CNN Prediction}\\ \cline{2-6}
   &  \centering 0 & \centering 1 & \centering 2 & \centering 3 & \centering 4 \cr \hline
  \centering 0 & \centering 90.51 \% & \centering 9.06 \% & \centering 0.33 \% & 0.10 \% & 0.00 \% \cr \hline
  \centering 1 & \centering 1.17 \% & \centering 79.42 \% & \centering 18.87 \% & 0.54 \% & 0.01 \% \cr \hline
  \centering 2 & \centering 0.07 \% & \centering 8.50 \% & \centering 78.92 \% & 12.42 \% & 0.09 \% \cr \hline
  \centering 3 & \centering 0.00 \% & \centering 0.82 \% & \centering 31.23 \% & 66.29 \% & 1.65 \% \cr \hline
  \centering 4 & \centering 0.00 \% & \centering 0.13 \% & \centering 5.51 \% & 76.14 \% & 18.22 \% \cr \hline
\end{tabular}} \caption{}
  \end{subtable}\par\bigskip
  \begin{subtable}[t]{.5\linewidth}%
    \centering%
\resizebox{\columnwidth}{!}{
\begin{tabular}{|p{12mm}|c|c|c|c|c|}
\hline
   \centering No. Sources &  \multicolumn{5}{|c|}{\vspace{-5mm}CNN Prediction}\\ \cline{2-6}
   &  \centering 0 & \centering 1 & \centering 2 & \centering 3 & \centering 4 \cr \hline
  \centering 0 & \centering 93.70 \% & \centering 6.16 \% & \centering 0.12 \% & 0.03 \% & 0.00 \% \cr \hline
  \centering 1 & \centering 0.93 \% & \centering 79.93 \% & \centering 18.95 \% & 0.19 \% & 0.00 \% \cr \hline
  \centering 2 & \centering 0.05 \% & \centering 7.29 \% & \centering 80.41 \% & 12.21 \% & 0.03 \% \cr \hline
  \centering 3 & \centering 0.00 \% & \centering 0.75 \% & \centering 26.55 \% & 71.14 \% & 1.56 \% \cr \hline
  \centering 4 & \centering 0.00 \% & \centering 0.12 \% & \centering 5.18 \% & 73.48 \% & 21.23 \% \cr \hline
\end{tabular}} \caption{}
  \end{subtable}
  \begin{subtable}[t]{.5\linewidth}%
    \centering%
    \resizebox{\columnwidth}{!}{
\begin{tabular}{|p{12mm}|c|c|c|c|c|}
\hline
   \centering No. Sources &  \multicolumn{5}{|c|}{\vspace{-5mm}CNN Prediction}\\ \cline{2-6}
   &  \centering 0 & \centering 1 & \centering 2 & \centering 3 & \centering 4 \cr \hline
  \centering 0 & \centering 95.17 \% & \centering 4.80 \% & \centering 0.03 \% & 0.00 \% & 0.00 \% \cr \hline
  \centering 1 & \centering 0.88 \% & \centering 78.23 \% & \centering 20.81 \% & 0.07 \% & 0.00 \% \cr \hline
  \centering 2 & \centering 0.05 \% & \centering 5.52 \% & \centering 79.87 \% & 14.54 \% & 0.02 \% \cr \hline
  \centering 3 & \centering 0.02 \% & \centering 0.49 \% & \centering 24.45 \% & 72.89 \% & 2.15 \% \cr \hline
  \centering 4 & \centering 0.03 \% & \centering 0.16 \% & \centering 4.03 \% & 73.22 \% & 32.57 \% \cr \hline
\end{tabular}} \caption{}
  \end{subtable}\par\bigskip
  \begin{subtable}[t]{.5\linewidth}%
    \centering%
\resizebox{\columnwidth}{!}{
\begin{tabular}{|p{12mm}|c|c|c|c|c|}
\hline
   \centering No. Sources &  \multicolumn{5}{|c|}{\vspace{-5mm}CNN Prediction}\\ \cline{2-6}
   &  \centering 0 & \centering 1 & \centering 2 & \centering 3 & \centering 4 \cr \hline
  \centering 0 & \centering 96.26 \% & \centering 3.74 \% & \centering 0.00 \% & 0.00 \% & 0.00 \% \cr \hline
  \centering 1 & \centering 0.82 \% & \centering 78.45 \% & \centering 20.66 \% & 0.07 \% & 0.00 \% \cr \hline
  \centering 2 & \centering 0.01 \% & \centering 4.46 \% & \centering 81.02 \% & 14.46 \% & 0.05 \% \cr \hline
  \centering 3 & \centering 0.00 \% & \centering 0.16 \% & \centering 20.71 \% & 76.68 \% & 2.45 \% \cr \hline
  \centering 4 & \centering 0.00 \% & \centering 0.00 \% & \centering 2.13 \% & 76.35 \% & 21.52 \% \cr \hline
\end{tabular}} \caption{}
  \end{subtable}
  \begin{subtable}[t]{.5\linewidth}%
    \centering%
    \resizebox{\columnwidth}{!}{
\begin{tabular}{|p{12mm}|c|c|c|c|c|}
\hline
   \centering No. Sources &  \multicolumn{5}{|c|}{\vspace{-5mm}CNN Prediction}\\ \cline{2-6}
   &  \centering 0 & \centering 1 & \centering 2 & \centering 3 & \centering 4 \cr \hline
  \centering 0 & \centering 91.39 \% & \centering 8.01 \% & \centering 0.43 \% & 0.14 \% & 0.04 \% \cr \hline
  \centering 1 & \centering 1.37 \% & \centering 78.75 \% & \centering 19.11 \% & 0.68 \% & 0.10 \% \cr \hline
  \centering 2 & \centering 0.10 \% & \centering 8.76 \% & \centering 77.82 \% & 13.05 \% & 0.27 \% \cr \hline
  \centering 3 & \centering 0.01 \% & \centering 0.87 \% & \centering 30.22 \% & 66.72 \% & 2.17 \% \cr \hline
  \centering 4 & \centering 0.01 \% & \centering 0.16 \% & \centering 5.86 \% & 74.60 \% & 19.37 \% \cr \hline
\end{tabular}} \caption{}
  \end{subtable}
  \caption{Source Number Prediction Performance at 50,000 background particle level. (a): One Linear Detector Array, (b): Two Adjacent Linear Detector Arrays, (c): Two Opposite Linear Detector Arrays, (d): Three Linear Detector Arrays, (e): Four Linear Detector Arrays, (f): Performance Across Whole Data Set (Every combination of detectors)}\label{tab:number_50k}
\end{table}

\begin{table}

  \begin{subtable}[t]{.5\linewidth}%
    \centering%
\resizebox{\columnwidth}{!}{
\begin{tabular}{|p{12mm}|c|c|c|c|c|}
\hline
   \centering No. Sources &  \multicolumn{5}{|c|}{\vspace{-5mm}CNN Prediction}\\ \cline{2-6}
   &  \centering 0 & \centering 1 & \centering 2 & \centering 3 & \centering 4 \cr \hline
  \centering 0 & \centering 31.21 \% & \centering 61.87 \% & \centering 5.37 \% & 1.29 \% & 0.26 \% \cr \hline
  \centering 1 & \centering 1.88 \% & \centering 60.02 \% & \centering 34.37 \% & 3.19 \% & 0.53 \% \cr \hline
  \centering 2 & \centering 0.36 \% & \centering 15.92 \% & \centering 69.30 \% & 13.38 \% & 1.04 \% \cr \hline
  \centering 3 & \centering 0.10 \% & \centering 3.15 \% & \centering 49.37 \% & 44.95 \% & 2.43 \% \cr \hline
  \centering 4 & \centering 0.00 \% & \centering 0.70 \% & \centering 19.63 \% & 71.29 \% & 8.37 \% \cr \hline
\end{tabular}} \caption{}
  \end{subtable}
  \begin{subtable}[t]{.5\linewidth}%
    \centering%
    \resizebox{\columnwidth}{!}{
\begin{tabular}{|p{12mm}|c|c|c|c|c|}
\hline
   \centering No. Sources &  \multicolumn{5}{|c|}{\vspace{-5mm}CNN Prediction}\\ \cline{2-6}
   &  \centering 0 & \centering 1 & \centering 2 & \centering 3 & \centering 4 \cr \hline
  \centering 0 & \centering 32.05 \% & \centering 64.82 \% & \centering 2.88 \% & 0.26 \% & 0.00 \% \cr \hline
  \centering 1 & \centering 1.02 \% & \centering 57.93 \% & \centering 39.83 \% & 1.19 \% & 0.03 \% \cr \hline
  \centering 2 & \centering 0.12 \% & \centering 9.84 \% & \centering 76.55 \% & 13.36 \% & 0.12 \% \cr \hline
  \centering 3 & \centering 0.01 \% & \centering 1.63 \% & \centering 41.18 \% & 56.22 \% & 0.96 \% \cr \hline
  \centering 4 & \centering 0.00 \% & \centering 0.35 \% & \centering 12.07 \% & 79.67 \% & 7.91 \% \cr \hline
\end{tabular}} \caption{}
  \end{subtable}\par\bigskip
  \begin{subtable}[t]{.5\linewidth}%
    \centering%
\resizebox{\columnwidth}{!}{
\begin{tabular}{|p{12mm}|c|c|c|c|c|}
\hline
   \centering No. Sources &  \multicolumn{5}{|c|}{\vspace{-5mm}CNN Prediction}\\ \cline{2-6}
   &  \centering 0 & \centering 1 & \centering 2 & \centering 3 & \centering 4 \cr \hline
  \centering 0 & \centering 34.00 \% & \centering 64.44 \% & \centering 1.44 \% & 0.09 \% & 0.03 \% \cr \hline
  \centering 1 & \centering 0.81 \% & \centering 57.32 \% & \centering 41.28 \% & 0.58 \% & 0.01 \% \cr \hline
  \centering 2 & \centering 0.10 \% & \centering 7.88 \% & \centering 78.02 \% & 13.92 \% & 0.09 \% \cr \hline
  \centering 3 & \centering 0.05 \% & \centering 1.22 \% & \centering 35.99 \% & 61.92 \% & 0.81 \% \cr \hline
  \centering 4 & \centering 0.00 \% & \centering 0.17 \% & \centering 10.47 \% & 80.58 \% & 8.77 \% \cr \hline
\end{tabular}} \caption{}
  \end{subtable}
  \begin{subtable}[t]{.5\linewidth}%
    \centering%
    \resizebox{\columnwidth}{!}{
\begin{tabular}{|p{12mm}|c|c|c|c|c|}
\hline
   \centering No. Sources &  \multicolumn{5}{|c|}{\vspace{-5mm}CNN Prediction}\\ \cline{2-6}
   &  \centering 0 & \centering 1 & \centering 2 & \centering 3 & \centering 4 \cr \hline
  \centering 0 & \centering 35.87 \% & \centering 63.64 \% & \centering 0.45 \% & 0.04 \% & 0.00 \% \cr \hline
  \centering 1 & \centering 0.65 \% & \centering 54.57 \% & \centering 44.47 \% & 0.31 \% & 0.00 \% \cr \hline
  \centering 2 & \centering 0.05 \% & \centering 6.00 \% & \centering 78.01 \% & 15.93 \% & 0.02 \% \cr \hline
  \centering 3 & \centering 0.01 \% & \centering 0.70 \% & \centering 32.67 \% & 65.65 \% & 0.96 \% \cr \hline
  \centering 4 & \centering 0.03 \% & \centering 0.17 \% & \centering 7.72 \% & 82.65 \% & 9.42 \% \cr \hline
\end{tabular}} \caption{}
  \end{subtable}\par\bigskip
  \begin{subtable}[t]{.5\linewidth}%
    \centering%
\resizebox{\columnwidth}{!}{
\begin{tabular}{|p{12mm}|c|c|c|c|c|}
\hline
   \centering No. Sources &  \multicolumn{5}{|c|}{\vspace{-5mm}CNN Prediction}\\ \cline{2-6}
   &  \centering 0 & \centering 1 & \centering 2 & \centering 3 & \centering 4 \cr \hline
  \centering 0 & \centering 35.85 \% & \centering 63.98 \% & \centering 0.17 \% & 0.00 \% & 0.00 \% \cr \hline
  \centering 1 & \centering 0.66 \% & \centering 55.64 \% & \centering 43.33 \% & 0.37 \% & 0.00 \% \cr \hline
  \centering 2 & \centering 0.00 \% & \centering 4.66 \% & \centering 79.39 \% & 15.93 \% & 0.02 \% \cr \hline
  \centering 3 & \centering 0.00 \% & \centering 0.32 \% & \centering 29.59 \% & 68.83 \% & 1.27 \% \cr \hline
  \centering 4 & \centering 0.00 \% & \centering 0.00 \% & \centering 4.60 \% & 84.58 \% & 10.82 \% \cr \hline
\end{tabular}} \caption{}
  \end{subtable}
  \begin{subtable}[t]{.5\linewidth}%
    \centering%
    \resizebox{\columnwidth}{!}{
\begin{tabular}{|p{12mm}|c|c|c|c|c|}
\hline
   \centering No. Sources &  \multicolumn{5}{|c|}{\vspace{-5mm}CNN Prediction}\\ \cline{2-6}
   &  \centering 0 & \centering 1 & \centering 2 & \centering 3 & \centering 4 \cr \hline
  \centering 0 & \centering 33.36 \% & \centering 63.61 \% & \centering 2.52 \% & 0.44 \% & 0.07 \% \cr \hline
  \centering 1 & \centering 1.10 \% & \centering 57.36 \% & \centering 40.04 \% & 1.35 \% & 0.15 \% \cr \hline
  \centering 2 & \centering 0.15 \% & \centering 9.83 \% & \centering 75.39 \% & 14.30 \% & 0.33 \% \cr \hline
  \centering 3 & \centering 0.04 \% & \centering 1.65 \% & \centering 39.63 \% & 57.33 \% & 1.35 \% \cr \hline
  \centering 4 & \centering 0.01 \% & \centering 0.35 \% & \centering 12.22 \% & 78.68 \% & 8.75 \% \cr \hline
\end{tabular}} \caption{}
  \end{subtable}
  \caption{Source Number Prediction Performance at 20,000 background particle level. (a): One Linear Detector Array, (b): Two Adjacent Linear Detector Arrays, (c): Two Opposite Linear Detector Arrays, (d): Three Linear Detector Arrays, (e): Four Linear Detector Arrays, (f): Performance Across Whole Data Set (Every combination of detectors)}\label{tab:number_20k}
\end{table}

\end{document}